\documentclass[12pt,preprint]{aastex}

\usepackage{lineno}
\usepackage{hyperref}
\usepackage{breakurl}

\begin{document}

\title{Search for Magnetically Broadened Cascade Emission From Blazars with VERITAS}


\author{
S.~Archambault\altaffilmark{1},
A.~Archer\altaffilmark{2},
W.~Benbow\altaffilmark{3},
M.~Buchovecky\altaffilmark{4},
V.~Bugaev\altaffilmark{2},
M.~Cerruti\altaffilmark{3},
M.~P.~Connolly\altaffilmark{5},
W.~Cui\altaffilmark{6,7},
A.~Falcone\altaffilmark{8},
M.~Fern\'{a}ndez Alonso\altaffilmark{9},
J.~P.~Finley\altaffilmark{6},
H.~Fleischhack\altaffilmark{10},
L.~Fortson\altaffilmark{11},
A.~Furniss\altaffilmark{12},
S.~Griffin\altaffilmark{1},
M.~H\"{u}tten\altaffilmark{10},
O.~Hervet\altaffilmark{13},
J.~Holder\altaffilmark{14},
T.~B.~Humensky\altaffilmark{15},
C.~A.~Johnson\altaffilmark{13},
P.~Kaaret\altaffilmark{16},
P.~Kar\altaffilmark{17},
D.~Kieda\altaffilmark{17},
M.~Krause\altaffilmark{10},
F.~Krennrich\altaffilmark{18},
M.~J.~Lang\altaffilmark{5},
T.~T.Y.~Lin\altaffilmark{1},
G.~Maier\altaffilmark{10},
S.~McArthur\altaffilmark{6},
P.~Moriarty\altaffilmark{5},
D.~Nieto\altaffilmark{15},
S.~O'Brien\altaffilmark{19},
R.~A.~Ong\altaffilmark{4},
A.~N.~Otte\altaffilmark{20},
M.~Pohl\altaffilmark{21,10},
A.~Popkow\altaffilmark{4},
E.~Pueschel\altaffilmark{19},
J.~Quinn\altaffilmark{19},
K.~Ragan\altaffilmark{1},
P.~T.~Reynolds\altaffilmark{22},
G.~T.~Richards\altaffilmark{20},
E.~Roache\altaffilmark{3},
A.~C.~Rovero\altaffilmark{9},
I.~Sadeh\altaffilmark{10},
K.~Shahinyan\altaffilmark{11},
D.~Staszak\altaffilmark{23},
I.~Telezhinsky\altaffilmark{21,10},
J.~Tyler\altaffilmark{1},
S.~P.~Wakely\altaffilmark{23},
A.~Weinstein\altaffilmark{18},
T.~Weisgarber\altaffilmark{24},
P.~Wilcox\altaffilmark{16},
A.~Wilhelm\altaffilmark{21,10},
D.~A.~Williams\altaffilmark{13},
B.~Zitzer\altaffilmark{1}
}
\altaffiltext{1}{Physics Department, McGill University, Montreal, QC H3A 2T8, Canada}
\altaffiltext{2}{Department of Physics, Washington University, St. Louis, MO 63130, USA}
\altaffiltext{3}{Fred Lawrence Whipple Observatory, Harvard-Smithsonian Center for Astrophysics, Amado, AZ 85645, USA}
\altaffiltext{4}{Department of Physics and Astronomy, University of California, Los Angeles, CA 90095, USA}
\altaffiltext{5}{School of Physics, National University of Ireland Galway, University Road, Galway, Ireland}
\altaffiltext{6}{Department of Physics and Astronomy, Purdue University, West Lafayette, IN 47907, USA}
\altaffiltext{7}{Department of Physics and Center for Astrophysics, Tsinghua University, Beijing 100084, China.}
\altaffiltext{8}{Department of Astronomy and Astrophysics, 525 Davey Lab, Pennsylvania State University, University Park, PA 16802, USA}
\altaffiltext{9}{Instituto de Astronom\'ia y F\'isica del Espacio (IAFE, CONICET-UBA), CC 67 - Suc. 28, (C1428ZAA) Ciudad Aut\'onoma de Buenos Aires, Argentina}
\altaffiltext{10}{DESY, Platanenallee 6, 15738 Zeuthen, Germany}
\altaffiltext{11}{School of Physics and Astronomy, University of Minnesota, Minneapolis, MN 55455, USA}
\altaffiltext{12}{Department of Physics, California State University - East Bay, Hayward, CA 94542, USA}
\altaffiltext{13}{Santa Cruz Institute for Particle Physics and Department of Physics, University of California, Santa Cruz, CA 95064, USA}
\altaffiltext{14}{Department of Physics and Astronomy and the Bartol Research Institute, University of Delaware, Newark, DE 19716, USA}
\altaffiltext{15}{Physics Department, Columbia University, New York, NY 10027, USA}
\altaffiltext{16}{Department of Physics and Astronomy, University of Iowa, Van Allen Hall, Iowa City, IA 52242, USA}
\altaffiltext{17}{Department of Physics and Astronomy, University of Utah, Salt Lake City, UT 84112, USA}
\altaffiltext{18}{Department of Physics and Astronomy, Iowa State University, Ames, IA 50011, USA}
\altaffiltext{19}{School of Physics, University College Dublin, Belfield, Dublin 4, Ireland}
\altaffiltext{20}{School of Physics and Center for Relativistic Astrophysics, Georgia Institute of Technology, 837 State Street NW, Atlanta, GA 30332-0430}
\altaffiltext{21}{Institute of Physics and Astronomy, University of Potsdam, 14476 Potsdam-Golm, Germany}
\altaffiltext{22}{Department of Physical Sciences, Cork Institute of Technology, Bishopstown, Cork, Ireland}
\altaffiltext{23}{Enrico Fermi Institute, University of Chicago, Chicago, IL 60637, USA}
\altaffiltext{24}{Department of Physics, University of Wisconsin-Madison, Madison, WI, 53706, USA}



\begin{abstract}
We present a search for magnetically broadened gamma-ray emission around active galactic nuclei (AGN), using VERITAS observations of seven hard-spectrum blazars. A cascade process occurs when multi-TeV gamma rays from AGN interact with extragalactic background light (EBL) photons to produce electron-positron pairs, which then interact with cosmic microwave background (CMB) photons via inverse-Compton scattering to produce gamma rays. Due to the deflection of the electron-positron pairs, a non-zero intergalactic magnetic field (IGMF) would potentially produce detectable effects on the angular distribution of the cascade emission. In particular, an angular broadening compared to the unscattered emission could occur. Through non-detection of angularly broadened emission from 1ES 1218+304, the source with the largest predicted cascade fraction, we exclude a range of IGMF strengths around $10^{-14}$~G at the 95\% confidence level. The extent of the exclusion range varies with the assumptions made about the intrinsic spectrum of 1ES~1218+304 and the EBL model used in the simulation of the cascade process. All of the sources are used to set limits on the flux due to extended emission.
\end{abstract}




\section{Introduction}
The intergalactic magnetic field (IGMF) is a postulated weak magnetic field permeating the voids between cosmological filaments. It provides a plausible seed field for the strong magnetic fields observed in galaxies and galaxy clusters, and is thus relevant for developing a complete picture of large-scale structure formation (see~\cite{Durrer2013} for a review). Charged cosmic rays will be deflected by the IGMF, complicating attempts to search for correlations between observed ultra-high-energy cosmic rays and potential extragalactic sources, such as active galactic nuclei (AGN) (see e.g. \cite{Sigl2004}).

A number of mechanisms have been discussed for the generation of the IGMF (see~\cite{Durrer2013} for a review). The field could be generated in the early universe during the epoch of inflation, the electroweak phase transition, or during recombination~\citep{Grasso2001}. A non-primordial IGMF could be generated via injection of magnetized plasma into the intergalactic medium by galactic winds~\citep{Bertone2006}. Many of these scenarios result in specific predictions for the IGMF strength and correlation length (the maximum length over which the magnetic field can be treated as coherent) that distinguish them from alternative models, although degeneracies exist for some combinations of strength and correlation length. Precise constraints on the parameters of the IGMF are needed to constrain models of the field's generation.

Though the strength and correlation length of the IGMF are constrained observationally, a broad swath of values for these quantities remains both theoretically and observationally allowed. Upper limits on the strength of the IGMF have been set with three methods: Zeeman splitting measurements of spectral lines, and Faraday rotation measurements of distant quasars and of the cosmic microwave background (CMB)~\citep{Heiles2004, Blasi1999, Polarbear2015, Planck2015}. For correlation lengths of 1~Mpc, the upper limits are on the order of 10$^{-9}$~G. Lower limits on the IGMF strength have been set based on studies of the gamma-ray emission from distant AGN, described in more detail below. Assuming correlation lengths of at least 1~Mpc, field strengths below about 10$^{-19}$~G have been excluded~\citep{Finke2015}, in addition to a small range of field strengths around 10$^{-15}$~G~\citep{HESS2014}.

Observations of AGN provide a means to probe the IGMF strength across much of the allowed range. Over 50 AGN are detected in the very-high-energy (VHE; $>$ 100 GeV) gamma-ray range\footnote{http://tevcat.uchicago.edu}, most of which are high-frequency-peaked BL Lacertae objects (HBLs). The high-energy photons emitted by these sources can be used as probes of the IGMF via their interactions with the extragalactic background light (EBL, see e.g.~\cite{Krennrich2013}) en route to the observer. As multi-TeV photons travel to the observer, they interact with EBL photons and produce electron-positron pairs. The trajectory of the electrons and positrons will be bent by the IGMF, extending the path length of the cascade emission with respect to the unscattered primary emission. The mean free path for the electrons and positrons is on the order of 10 kpc for the energies and redshifts considered in the study described below~\citep{Aharonian2009}. The electrons/positrons eventually up-scatter low-energy CMB photons to GeV energies via inverse-Compton scattering. The GeV photons can again pair-produce on the EBL, leading to an electromagnetic cascade. Due to the deflection of the electrons and positrons before inverse-Compton scattering, the cascade emission will be angularly broadened and time delayed~\citep{Aharonian1994, Plaga1995}. The time delay varies from hours to years, depending on the energy of the cascade photons and the IGMF strength. Furthermore, the cascade emission will have a lower average energy than the primary emission.

The energy distribution and angular and temporal properties of the cascade emission provide observable signatures, but measurements are challenging, given the limited sensitive energy range of existing gamma-ray instruments. Several efforts have been made to constrain the IGMF based on the shape of spectral energy distributions of AGN in the GeV to the TeV energy range~\citep{Taylor2011, Arlen2014, Finke2015}, using spectral measurements by $Fermi$-LAT below $\sim$100~GeV and imaging atmospheric Cherenkov telescope (IACT) arrays above $\sim$100~GeV. The interpretation of results is complicated by the different sensitivities of the instruments used in the measurements and the use of data from non-contemporaneous observations of variable sources.

The currently operating IACT arrays can be used independently to search for an IGMF-induced angular broadening of cascade emission. However, to produce cascade emission above the energy thresholds of these instruments, the initial photons must have multi-TeV energies (see Eq. 27 of~\cite{Neronov2009} for an approximate relation between the initial and cascade photon energies). The best candidates for searches for cascade emission with IACTs are thus extreme-HBLs, whose spectral energy distributions exhibit a high-frequency peak at $\sim$1~TeV and hard spectral indices~\citep{Bonnoli2015}. For several of these sources, no evidence of a spectral break/high-energy cutoff is observed in the intrinsic energy spectrum. It is possible that the primary emission follows an unbroken power-law distribution to several tens of TeV for these sources, as will be discussed in Section~\ref{sourceselsec}.  

The magnitude of the angular broadening and the time delay varies with the IGMF strength and can be divided into three regimes. For 10$^{-12}$~G~$\lesssim$~B~$<$~10$^{-7}$~G, the electron-positron pairs will be isotropized in the vicinity of the blazar, forming a pair halo that manifests itself as a broader spatial emission than expected for a point source~\citep{Aharonian2009}. As the mean free path before production of the electron-positron pairs does not depend on the magnetic field strength, neither will the size of the predicted pair halo. However, as a stronger field will isotropize higher energy electron-positron pairs, in the event of a pair halo detection the energy distribution of the cascade emission will provide information about the field strength. 

For a field strength of 10$^{-16}$~G~$<$~B~$\lesssim$~10$^{-12}$~G, the bulk of the electron-positron pairs do not isotropize, but as in the pair halo regime, the cascade  produces an angularly broadened emission component in addition to the unscattered emission~\citep{Elyiv2009}. The magnitude of the broadening is proportional to the field strength.

For B~$<10^{-16}$~G, the predicted angular broadening is too small to be resolved by currently operating IACTs, and in the VHE range the cascade can only be detected via the observation of a time delayed component following a source flare~\citep{Plaga1995, Dermer2011} or by the observation of angularly broadened emission in the GeV band. The study here focuses on the search for angular broadening rather than time delays, and is thus insensitive to field strengths of B~$<10^{-16}$~G. 

In addition to the dependence on the intrinsic source spectrum, the projected sensitivity to angularly broadened cascade emission depends on the source redshift, including the evolution of the EBL with redshift. At redshifts $z$~$\gtrsim$~0.2, the EBL intensity is high and results in a short mean free path for both gamma rays and electron-positron pairs, producing cascade emission that is not easily distinguished from the primary emission in a spatial analysis. In contrast, for nearby ($z$~$\lesssim$~0.1) sources, the cascade emission is too broad to be easily distinguishable from the isotropic cosmic-ray background~\citep{Aharonian2009}. Additionally, the distance between the source and the first pair production interaction typically exceeds the distance between the source and Earth for sources that are $\sim$100 Mpc away, resulting in a small predicted cascade fraction.

Previous searches for angularly broadened emission around blazars have been performed by MAGIC using Mrk 501 and Mrk 421~\citep{MAGIC2010}, $Fermi$--LAT using a large blazar sample~\citep{FermiLAT2013}, VERITAS using Mrk 421~\citep{MateoProc} and H.E.S.S. using 1ES~1101-232, 1ES~0229+200 and PKS~2155-304~\citep{HESS2014}. All searches resulted in non-detections, with H.E.S.S. excluding IGMF strengths of (0.3--3)$\times$10$^{-15}$~G at the 99\% confidence level. \cite{Chen2015} claim a pair halo detection based on $Fermi$--LAT blazar observations, corresponding to an IGMF strength of 10$^{-17}$--10$^{-15}$~G. Although not confirmed, the suggested range partially overlaps the expected sensitivity range for IACT searches for angularly broadened emission.

It has been questioned whether cascade emission can be expected to reach the observer, as it is possible that the energy of the cascade could be entirely dissipated by collective behavior of the charged particles in the cascade~\citep{Broderick2012}. In this scenario, energy losses due to plasma beam instabilities would dominate over the cooling of the electron-positron pairs by inverse-Compton scattering. This proposal has been argued against elsewhere in the literature, and thus the impact of plasma instabilities remains an open question~\citep{Schlickeiser2012, Miniati2013}. Plasma instabilities are not considered in the cascade simulations used in this study.

\section{Observations}
VERITAS is an array of four 12 meter IACTs \citep{VERITASinstrument} located at the Fred Lawrence Whipple Observatory in southern Arizona, USA (+31$^{\circ}$ 40$\arcmin$30.21$\arcsec$, --110$^{\circ}$ 57$\arcmin$ 7.77$\arcsec$, 1268 m above sea level). The four telescopes are of Davies-Cotton design~\citep{Davies1957}. The VERITAS cameras are each instrumented with 499 photomultiplier tubes (PMTs). Dead space between the PMTs is minimized with the use of light cones. VERITAS detects Cherenkov emission induced by particle showers in the atmosphere and is sensitive to gamma rays with energies from $\sim$85 GeV up to greater than 30 TeV. One of the telescopes was moved in 2009 to create a more symmetric array, improving the instrument sensitivity. A major camera upgrade was completed in 2012, which decreased the lower bound on the energy range from $\sim$100 GeV~\citep{Kieda2013}. The instrument has a field of view of 3.5$^{\circ}$, an energy resolution of 15--25\%, and an angular resolution (given as the 68\% containment radius) of $<$0.1$^{\circ}$ at 1 TeV~\citep{VERITASspecs}. 

Data were collected in \textit{wobble} pointing mode, with the camera center offset by 0.5$^{\circ}$ from the source position, allowing for the simultaneous collection of data from the source and background regions within the same field of view~\citep{Fomin}.

The data used in this analysis were collected between 2009 and 2012, after the array upgrade but before the camera upgrade. Obtaining the best possible angular resolution and the lowest possible energy threshold is critical for this measurement. Consequently, the data sample was restricted to observations made with zenith angles $<$30$^{\circ}$ and all four telescopes operating. However, characterization of the instrument performance is equally critical, thus a mature dataset was used for this work. A study of observations taken after the camera upgrade will be considered in a future publication. 

\section{Source Selection}\label{sourceselsec}
The sources used in this analysis were selected for optimal sensitivity to magnetically broadened emission. The amount of cascade emission falling within the sensitive energy range of VERITAS is greatest for hard-spectrum sources with intrinsic emission to $\sim$10 TeV. The intrinsic emission in the VHE band is attenuated by interaction with the EBL, resulting in an observed spectral index that is softer than the intrinsic value. Attenuation in the high-energy (HE; $<$ 100 GeV) range is expected to be minimal. Thus, spectral indices measured by $Fermi$--LAT are used as proxies for the intrinsic source indices, with values taken from the 3LAC catalog~\citep{3LAC}. Sources with indices harder than $\sim$2 were selected. 

To minimize statistical uncertainties, only sources with a detection significance of $>$~7.5$\sigma$ significance (calculated using Equation 17 of \cite{LiMa}) were selected. Although sources in this optimal range of redshifts were preferentially selected, sources at higher and lower redshifts were also included in the analysis. In the case of a detection of angularly broadened emission, this would allow tests of the redshift dependence of the broadening. An angular broadening should not be detectable for the most distant and closest sources, and the detection of a spatial extension would suggest an underestimated gamma-ray point spread function. 

A test was performed for an energy cutoff/spectral break below the highest-energy spectral point measured by VERITAS. The observed VHE spectra were corrected for EBL absorption with the fiducial model of \cite{Gilmore2012}. This model was selected based on its consistency with observational constraints on the EBL intensity~\citep{Biteau2015}. Each EBL-deabsorbed spectrum was fit with a power law, and the spectral index compared to the $Fermi$--LAT index. If the two indices disagreed by more than 1$\sigma$, this was considered an indication of a spectral break or exponential cutoff within the energy range covered by the two spectra (note that this results in a conservative prediction of the cascade emission fraction). For PG 1553+113, the source redshift is not well constrained, making it difficult to estimate the impact of the EBL absorption on the intrinsic VHE spectrum and consequently to check for the presence of a spectral break. Consequently, only limits on the flux due to extended emission are extracted for PG 1553+113, which does not require an assumption about the energy cutoff.

During strong flares, it is expected that the primary emission will be significantly brighter than the cascade emission, decreasing the overall sensitivity to the cascade component. Furthermore, rapid spectral variability can occur during flares, increasing the uncertainty on the predicted fraction of cascade emission. Thus, periods of high source activity were removed from the datasets of the highly variable sources Mrk~421 and Mrk~501 (observations with integral fluxes below 9$\times$10$^{-10}$~cm$^{-2}$s$^{-1}$ and 2$\times$10$^{-10}$~cm$^{-2}$s$^{-1}$ above 200~GeV, respectively, were retained). However, the high flux data were used to verify the point source simulation procedure described in Section~\ref{primarysimsec}. The remaining sources did not show significant flux variability within the selected datasets. The impact of spectral variability is further addressed in Section~\ref{IGMFconstraints}.

The final source list is shown in Table~\ref{sourcelist}, with the assumed intrinsic spectral index, indication for a spectral cutoff or break, redshift, and detection significance in the VERITAS data sample used in this study. For the assumptions about the intrinsic spectrum and EBL intensity taken, the predicted cascade fraction is below the VERITAS sensitivity for all of the sources other than 1ES 1218+304. However, these assumptions are subject to large uncertainties which strongly affect the predicted cascade fraction, as will be shown in Section\ref{IGMFconstraints}. Consequently, although only 1ES 1218+304 will be used to set limits on the cascade fraction and IGMF strength, the other sources are included in the analysis, in light of the possibility that the cascade fraction could be underpredicted for the nominal set of assumptions. Limits on the flux due to extended emission will be derived for all the sources.

\begin{table}
\centerline{
\begin{tabular}{cccccccc}
Source name & $z$ & $\Gamma$ & Cutoff & $T$~[min] & $N_{excess}$ & $\sigma_{detect}$ & $p-value$ \\
\hline
\hline
Mrk 421 & 0.031 & 1.772$\pm$0.008 & Y & 2269 & 21388 & 185.3 & 0.19\\
Mrk 501 & 0.034 & 1.716$\pm$0.016 & Y & 1389 & 7339 & 94.8 & 0.38\\
VER J0521+211 & 0.108 & 1.923$\pm$0.024 & Y & 990 & 649 &  23.2 & 0.31\\
H 1426+428 & 0.129 & 1.575$\pm$0.085 & Y & 1586 & 659 & 7.6 & 0.95\\
1ES 0229+200 & 0.139 & 2.025$\pm$0.150 & N & 3634 & 810 & 10.3 & 0.30\\
1ES 1218+304 & 0.182 &  1.660$\pm$0.038 & N & 3481 & 3420 & 35.5 & 0.52\\
PG 1553+113 & 0.4-0.6 & 1.604$\pm$0.025 & ? & 4502 & 4852 & 46.0 & 0.003\\
\hline
\hline
\end{tabular}
}
\caption{\small{Source properties. Column 1: Source name. Column 2: Redshift. Column 3: Assumed intrinsic spectral index (given by the $Fermi$--LAT measured index~\citep{3LAC}). Column 4: Indication of presence of a intrinsic spectral cutoff or break. Column 5: Exposure time. Column 6: Number of excess events. Column 7: VERITAS detection significance. Column 8: Probability that the $\theta^{2}$ histograms in data and simulation are drawn from the same distribution.}}
\label{sourcelist}
\end{table}

\section{Data Analysis}
All data were processed with the standard VERITAS calibration and shower parameterization pipelines~\citep{MDICRC}. To achieve the best possible angular resolution, events with shower images recorded in only two telescopes were discarded, while events with three or four images were retained. Gamma-hadron separation was achieved using selection on Hillas parameters~\citep{Hillas, Weekes1989}, with information from multiple images combined into the mean reduced scaled width and length as described by \cite{MDICRC}. For the sources Mrk~421, Mrk~501, 1ES~1218+304 and PG~1553+113, gamma-ray events were selected with box cuts on the individual gamma-hadron separation parameters. The event selection was optimized for a gamma-ray source with a soft spectral index ($\Gamma$~$>$~3.5), thus maximizing the sensitivity to low-energy cascade emission. 

The remaining sources---H~1426+428, 1ES 0229+200, and VER~J0521+211---were not detected with high significance using the standard gamma-hadron separation described above. For these sources a boosted decision tree (BDT) analysis was used, resulting in substantial improvements in the source detection significances~\citep{BDTpaper}. The BDT analysis incorporates the same gamma-hadron separation variables used in the standard analysis into a single discriminator. 

The detection significances quoted here differ from previously published VERITAS results, due to differences in the data samples and event selection. The results on VERITAS and multiwavelength observations on H~1426+428 will be presented in a separate paper (in preparation).

Both the data analysis and the simulations described below are restricted to an energy range from 160 GeV to 1 TeV. The lower limit is motivated by the energy threshold after analysis cuts (defined as the energy above which the energy bias falls below 10\%). The upper limit is imposed to minimize the systematic uncertainties associated with high-energy reconstruction, while retaining sensitivity to the low-energy cascade emission.
 
\section{Simulation of the Point Source Emission}\label{primarysimsec}
The distribution of the parameter $\theta^{2}$, defined as the squared angular distance between an air shower's reconstructed arrival direction and the target's estimated location, characterizes the angular profile of a source. A straightforward test for the presence of angularly broadened emission is to compare the measured $\theta^{2}$ distribution of a source against that of a known point source. For a point source, the width of the $\theta^{2}$ distribution is due only to the instrument's point spread function (PSF). 

Each point source was modeled with the standard VERITAS Monte Carlo simulation pipeline, which uses the CORSIKA program~\citep{CORSIKA} to model the interaction of gamma rays in the atmosphere and the GrISU package\footnote{http://www.physics.utah.edu/gammaray/GrISU/} to model the detector response. The VERITAS gamma-ray PSF depends on the reconstructed gamma-ray energy, the zenith and azimuthal angles of observation (as the PSF is impacted by the Earth's geomagnetic field), and the night sky background level. Consequently, the Monte Carlo simulations were weighted event-by-event to match their data counterparts' energy and azimuthal angular distributions. The range and mean value of the simulated night sky background level were matched to observations. The simulated sources were generated assuming a zenith angle of observations of 20$^{\circ}$, which only approximates the zenith-angle distributions of the data. To correct for the simplification, a function PSF($Ze$) (where $Ze$ is the zenith angle of observations) was derived from a large sample of Crab Nebula observations. With this function, the width $w_{obs}$ corresponding to the observed zenith-angle distribution was calculated. The difference ($w_{obs} - w_{20^{\circ}}$) was used to correct the fitted width of the simulated sources. The magnitude of the zenith-angle corrections ranges from 0.0009$^{\circ}$ to 0.0023$^{\circ}$. 

The uncertainty in the telescope pointing of 25$''$~\citep{Griffiths2015} translates into a broadening of the $\theta^{2}$ distribution by 0.0005$^{\circ}$, which was determined by shifting the assumed source position in a large Crab Nebula data sample from its nominal value within the pointing uncertainty.
	
The uncertainty in the energy scale of VERITAS introduces a further systematic uncertainty on the $\theta^{2}$ distribution. An average energy-scale uncertainty of 20\% was assumed. The impact of the energy-scale uncertainty on the $\theta^{2}$ distributions varies depending on the source spectrum, and thus was calculated for each source separately. The energy distributions used to weight the simulation were shifted up and down by 20\%, and the change in width of the resulting $\theta^{2}$ distributions taken as the uncertainty. The resulting uncertainties are listed in Table~\ref{fitwidths}. The energy-scale uncertainty dominates the uncertainty in the simulations, exceeding the statistical errors and pointing uncertainty by a few to several factors.

It is worth noting that the systematic uncertainties are small in comparison with the width of the $\theta^{2}$ distributions, on order of several percent, indicating that the angular resolution of VERITAS is the limiting factor in the sensitivity to angularly broadened emission.

\section{Simulation of the Cascade Process}\label{cascadesimsec}
Predictions for the angular profiles of the magnetically broadened cascade emission were calculated via a dedicated Monte Carlo simulation.
This simulation is based on the code presented in~\cite{WeisgarberPhD} but includes substantial modifications improving both the speed and accuracy of the calculations.
The code tracks the three-dimensional trajectories of electrons, positrons, and gamma rays in a $\Lambda$CDM spacetime, and it employs the full relativistic cross sections for the pair-production and inverse-Compton interactions with the redshift-dependent CMB and EBL populations.
Complete kinematic modeling for both interactions enables accurate predictions for arbitrarily small IGMF strengths, while an assumption that the electron-positron pairs do not isotropize limits the code's range of validity to IGMF strengths below $\sim10^{-12}$~G. Thirteen IGMF strengths logarithmically spaced in the range B~=~10$^{-16}$--10$^{-13}$~G are considered in the following.

For each particle, the simulation also calculates the amount of time accumulated between the injection of the initial gamma ray and the particle's instantaneous position. This time is then compared to the amount of time that would have been accumulated by a non-interacting gamma ray propagating directly from the source to the same position. The difference between these two times is tracked by the simulation at all points along every particle's trajectory.

To model the IGMF, a simple redshift scaling ${\bf B}(z)=B_0(1+z)^2{\bf\hat{b}}$ is assumed, with $B_0$ fixed to a constant value throughout all space and the unit vector ${\bf\hat{b}}$ encapsulating the direction of the field. A cubic grid with sides of length 1 Mpc in comoving coordinates represents the correlation length of the IGMF. This value of the correlation length is experimentally allowed for a broad range of IGMF strengths~\citep{Taylor2011}, and has been used in similar studies (e.g. \cite{HESS2014}). Within a cube, the code selects a fixed random direction for ${\bf\hat{b}}$, independent of the directions in the neighboring cubes. 

For each source redshift, the code samples gamma-ray energies from a distribution uniform in logarithmic energy between 0.15 and 500 TeV, injects the gamma rays at the redshift of the source, and tracks the resulting cascades.
Particles are recorded once their comoving distance from the source is equal to the comoving distance between the source and Earth.
The cascades are thinned to improve the statistical independence of the results. Weights are applied to the recorded events to account for the cascade thinning and to allow predictions to be obtained for arbitrary intrinsic spectra. The simulated EBL-corrected spectra were compared to measured spectra for several test cases, and found to be in good agreement. The predictions are valid for spectra that cut off at observed energies much lower than $500/(1+z)$ TeV. The bulk Lorentz factor and the viewing angle of the blazar jet were set as 10 and $0^{\circ}$, respectively.

\section{Results}
\begin{figure}
\centerline{\includegraphics[width=0.5\textwidth]{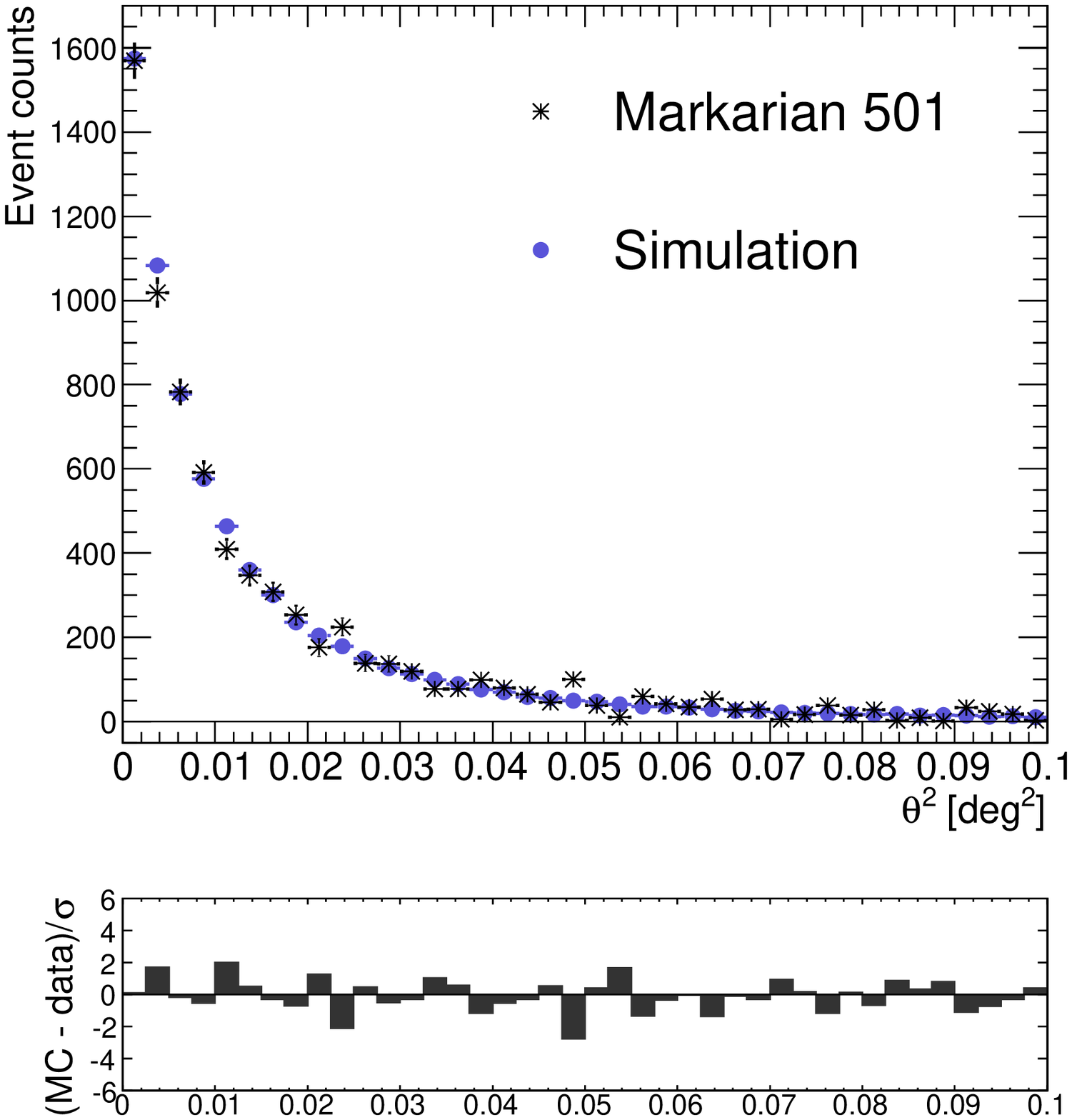}
\includegraphics[width=0.5\textwidth]{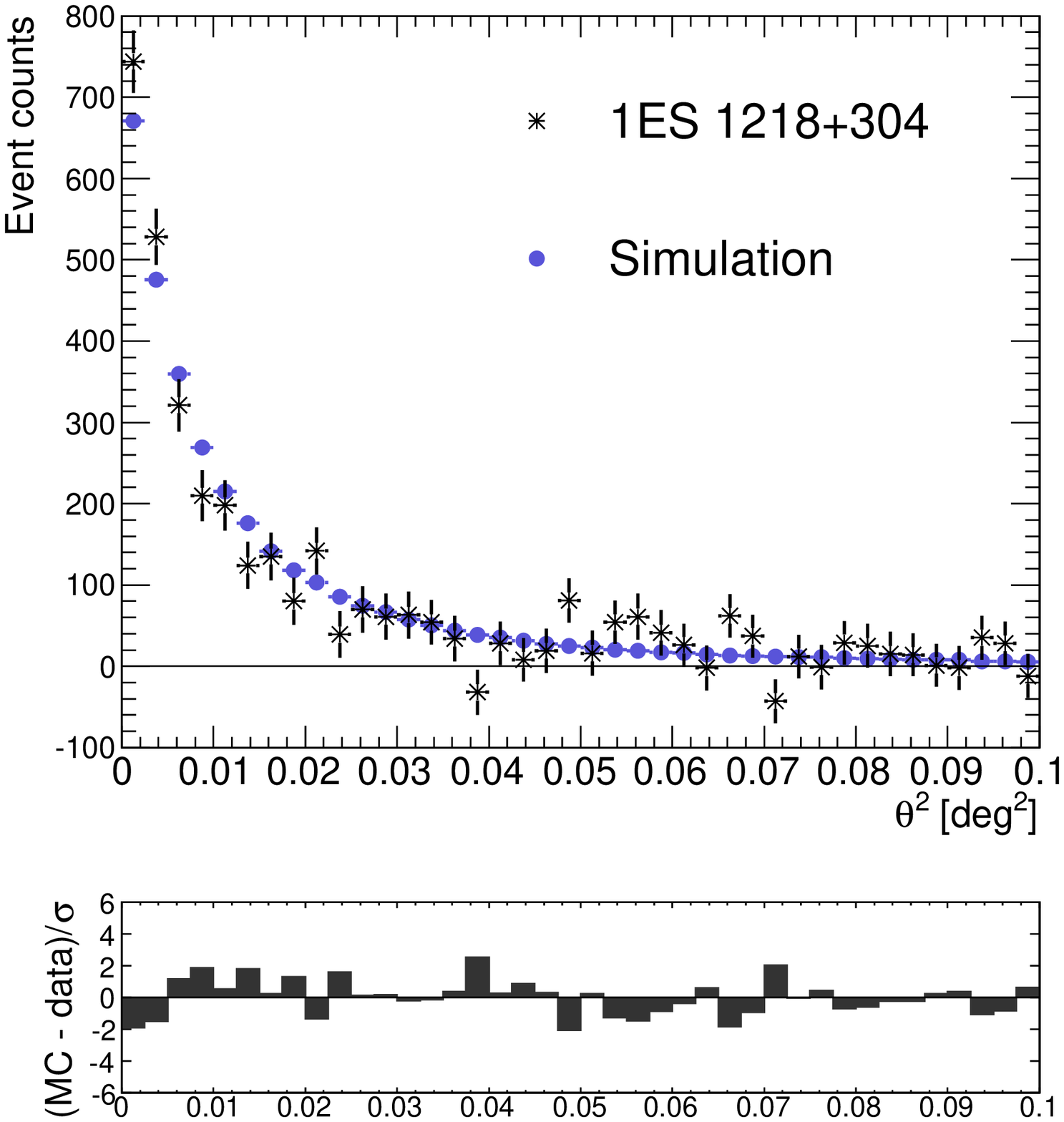}}
\caption{\small{Comparison between the angular profiles of Mrk~501 and 1ES~1218+304 and their simulated counterparts. The results of a $\chi^{2}$ probability test are shown in Table~\protect\ref{sourcelist} for all sources.}}\label{dataMCoverlay}
\end{figure}

The agreement between the measured and simulated $\theta^{2}$ distributions was first assessed based on the derived residual distributions and a $\chi^{2}$ probability test. Figure~\ref{dataMCoverlay} shows the $\theta^{2}$ distributions for Mrk~501 and 1ES~1218+304 and their simulated counterparts. The distributions are plotted for $\theta^{2}$=0.00--0.10~deg$^{2}$ in order to show detail, but the $\chi^{2}$ probability test is applied to the residual distributions between $\theta^{2}$=0.00~deg$^{2}$ and $\theta^{2}$=0.24~deg$^{2}$. The upper boundary on the $\theta^{2}$ range was selected based on the predicted width of the $\theta^{2}$ distribution when including a cascade component. The distributions for the remaining sources are given in Appendix~\ref{appendixA}. Note that the systematic uncertainties and zenith-angle corrections are not accounted for in Fig.~\ref{dataMCoverlay}. Even ignoring these effects, the $\chi^{2}$ probability test indicates good agreement between data and simulations. The $p$-values from the $\chi^{2}$ probability tests are shown in Table~\ref{sourcelist}. Only for PG~1553+113 does the $p$-value indicate a mismatch between data and simulation, at the 3$\sigma$ level, a tension that is resolved when the systematic uncertainties and zenith correction are considered.

In order to include the systematic uncertainties and the zenith-angle correction in the comparison of the simulation and data, fits of the $\theta^{2}$ distributions were performed. The simulated $\theta^{2}$ distributions are well-described by a polynomial multiplied with a hyberbolic secant of width $w$~\citep{Zitzer2013}: 

\begin{equation}
\textrm{P}(\theta) = (c_{0} + c_{1}\theta^{2} + c_{2}\theta^{4})~\textrm{sech}(\theta/w).\label{eq:Zitzerfunc}
\end{equation}
In order to facilitate comparison, the data distributions were fit with the same function, but with the allowed range of the parameters $c_{n}$ restricted to the 1$\sigma$ uncertainty band on the fitted values from simulation, $c_{n}^{sim}\pm\sigma_{c_{n}^{sim}}$. Note that allowing a broader range for the parameters $c_{n}$ (i.e. the 2$\sigma$ uncertainty band) did not change the fitted values of $w$.

\begin{figure}
\centerline{\includegraphics[width=0.5\textwidth]{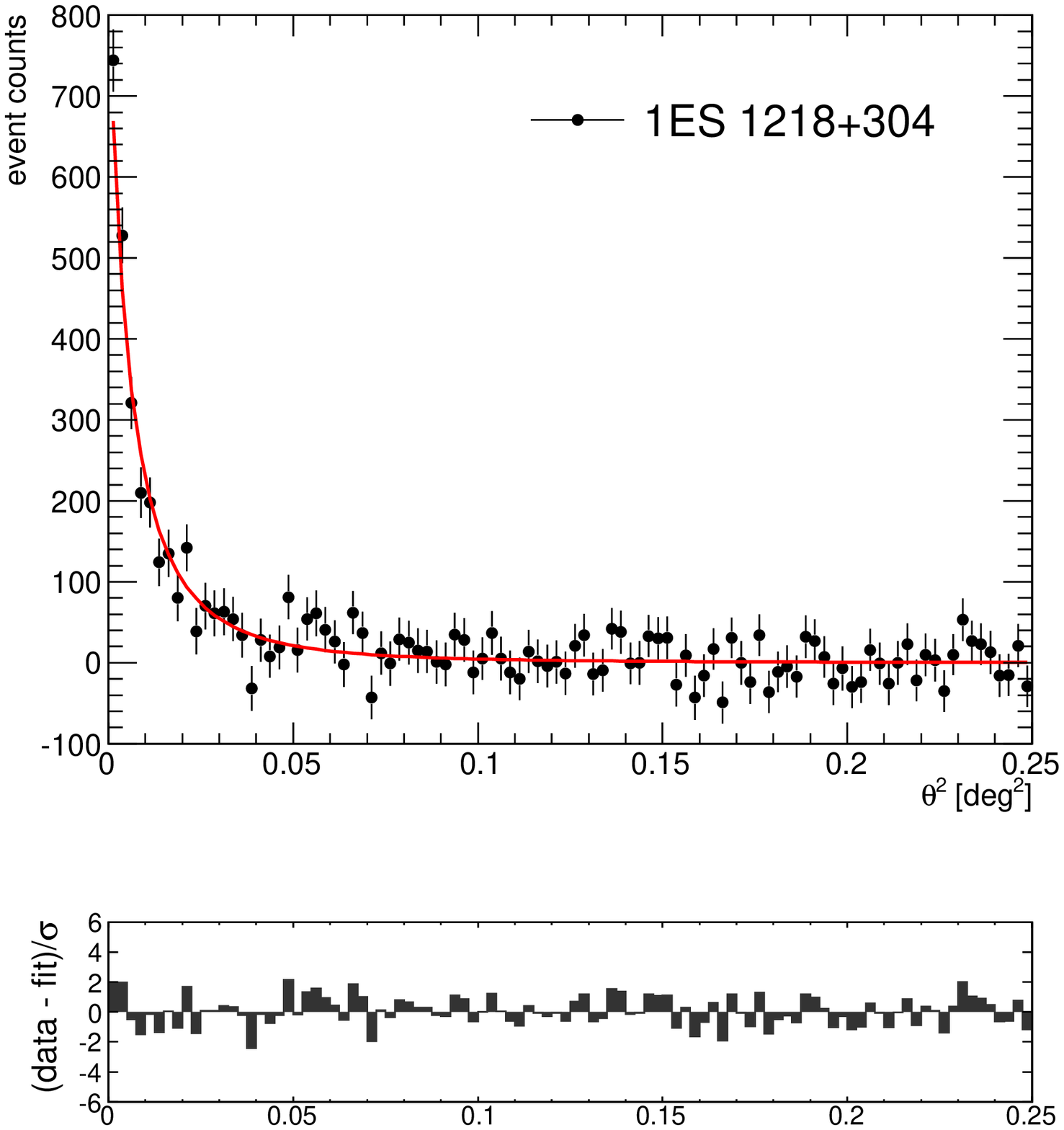}
\includegraphics[width=0.5\textwidth]{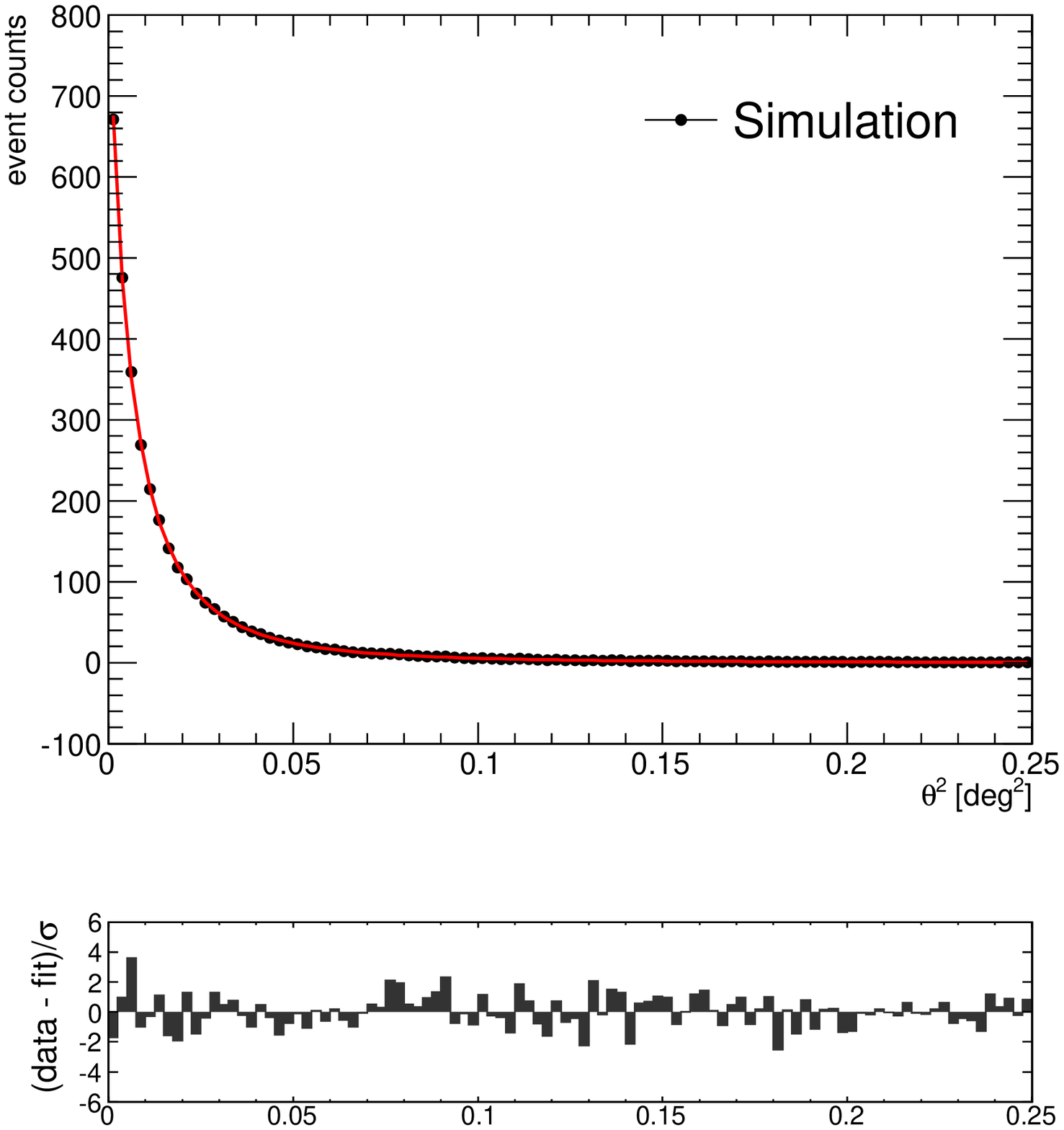}}
\caption{\small{Fitted $\theta^{2}$ distribution for 1ES 1218+304 and its simulated counterpart.}}\label{fit1ES1218}
\end{figure}

Fig.~\ref{fit1ES1218} shows the fitted $\theta^{2}$ distributions for data and simulation for 1ES 1218+304. The figures for the others sources are given in Appendix~\ref{appendixB}. The best-fit width parameters $w_{data}$ and $w_{sim}$ are compared in Table~\ref{fitwidths}. Note that for $w_{sim}$, the zenith-angle correction has been applied. The statistical and systematic uncertainties on the simulated widths are added in quadrature to produce the total uncertainty $\sigma_{sim}$. There is no significant discrepancy between data and simulations, nor are the data distributions systematically broader or narrower than the simulated distributions. The agreement is quantified in the figure of merit $s$=$(w_{data} - w_{sim})/\sqrt{(\sigma_{data}^{2} + \sigma_{sim}^{2})}$, shown in the final column of Table~\ref{fitwidths}.

\begin{table}[!htp]
\centerline{
\begin{tabular}{cccc}
Source name & $w_{data} \pm \sigma_{stat}^{data}$ & $w_{sim} \pm \sigma_{stat}^{sim} \pm \sigma_{pointing} \pm \sigma_{energy~scale}$  & $s$ \\
\hline
\hline
Mrk 421                & 0.0496$^{\circ}$ $\pm$ 0.0003$^{\circ}$ & 0.0484$^{\circ}$ $\pm$ 0.0002$^{\circ}$ $\pm$ 0.0005$^{\circ}$ $\pm$ 0.0017$^{\circ}$ & 0.7 \\
Mrk 501                & 0.0495$^{\circ}$ $\pm$ 0.0004$^{\circ}$ & 0.0481$^{\circ}$ $\pm$ 0.0003$^{\circ}$ $\pm$ 0.0005$^{\circ}$ $\pm$ 0.0011$^{\circ}$ & 1.1 \\
VER J0521+211 & 0.0477$^{\circ}$ $\pm$ 0.0019$^{\circ}$ & 0.0451$^{\circ}$ $\pm$ 0.0002$^{\circ}$ $\pm$ 0.0005$^{\circ}$ $\pm$ 0.0024$^{\circ}$ & 0.8 \\ 
H 1426+428        & 0.0447$^{\circ}$ $\pm$ 0.0047$^{\circ}$ & 0.0547$^{\circ}$ $\pm$ 0.0003$^{\circ}$ $\pm$ 0.0005$^{\circ}$ $\pm$ 0.0027$^{\circ}$ & -1.8 \\ 
1ES 0229+200   & 0.0395$^{\circ}$ $\pm$ 0.0040$^{\circ}$ & 0.0461$^{\circ}$ $\pm$ 0.0003$^{\circ}$ $\pm$ 0.0005$^{\circ}$ $\pm$ 0.0016$^{\circ}$ & -1.5 \\
1ES 1218+304   & 0.0512$^{\circ}$ $\pm$ 0.0012$^{\circ}$ & 0.0507$^{\circ}$ $\pm$ 0.0003$^{\circ}$ $\pm$ 0.0005$^{\circ}$ $\pm$ 0.0012$^{\circ}$ & 0.3 \\ 
PG 1553+113     & 0.0497$^{\circ}$ $\pm$ 0.0011$^{\circ}$ & 0.0521$^{\circ}$ $\pm$ 0.0002$^{\circ}$ $\pm$ 0.0005$^{\circ}$ $\pm$ 0.0022$^{\circ}$ & -1.0 \\ 
\hline
\hline
\end{tabular}
}
\caption{\small{Fitted width of the $\theta^{2}$ distribution for data and simulations. In the final column, $s$=$(w_{data} - w_{sim})/\sqrt{(\sigma_{data}^{2} + \sigma_{sim}^{2})}$}.}
\label{fitwidths}
\end{table}

\subsection{Limits on the IGMF Strength}\label{IGMFconstraints}
\begin{figure}
\centerline{\includegraphics[width=0.45\textwidth]{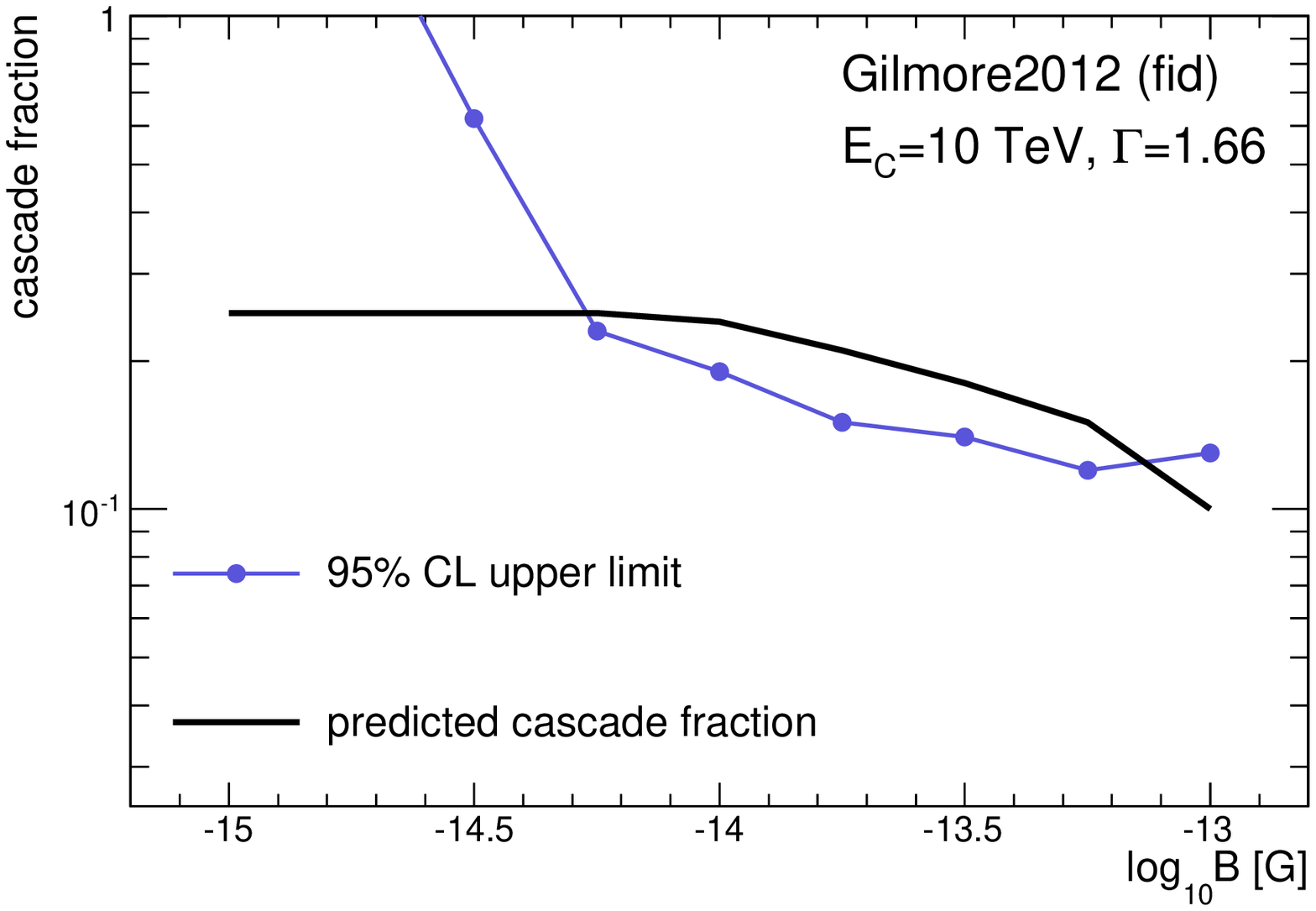}
\includegraphics[width=0.45\textwidth]{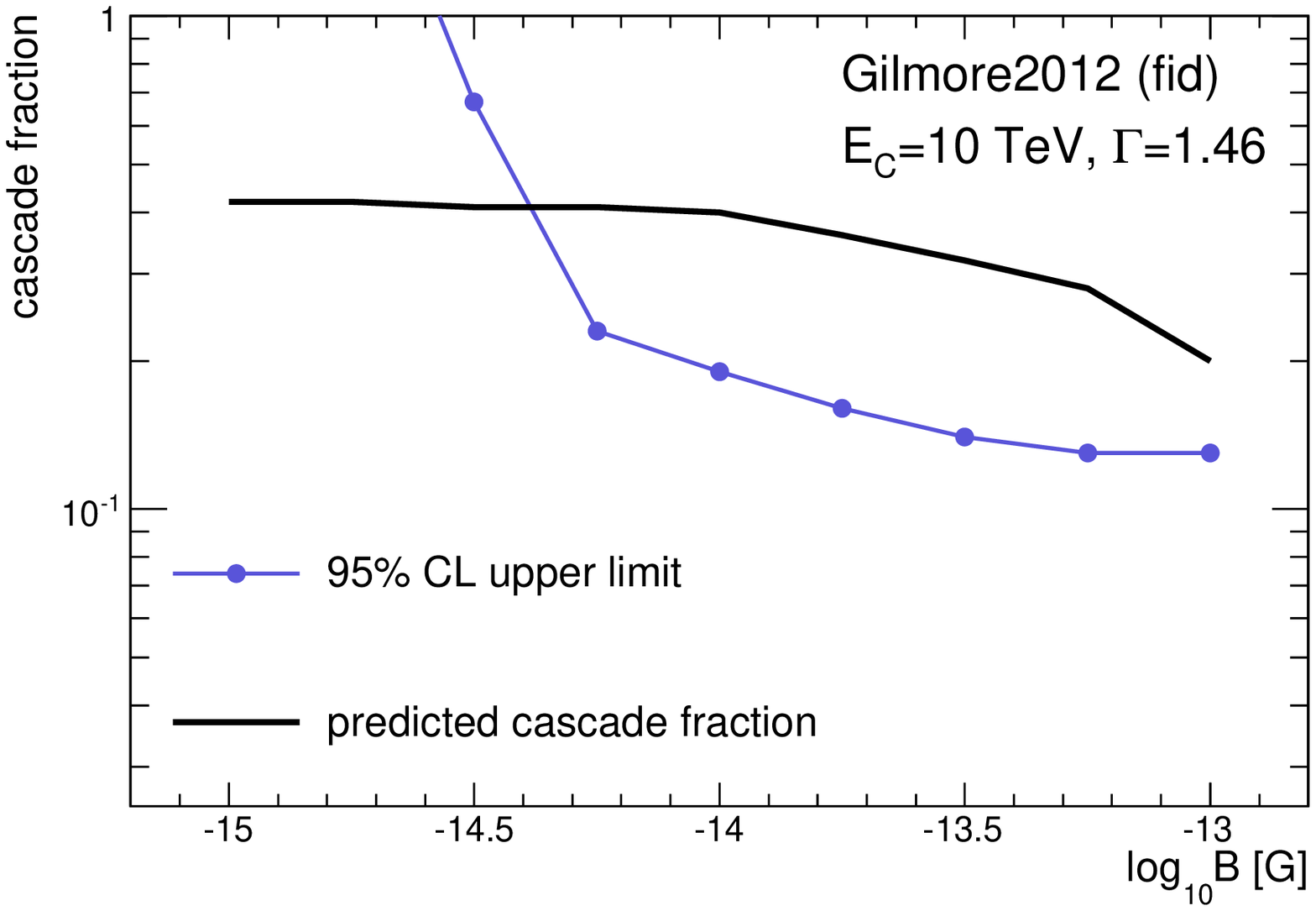}}
\centerline{\includegraphics[width=0.45\textwidth]{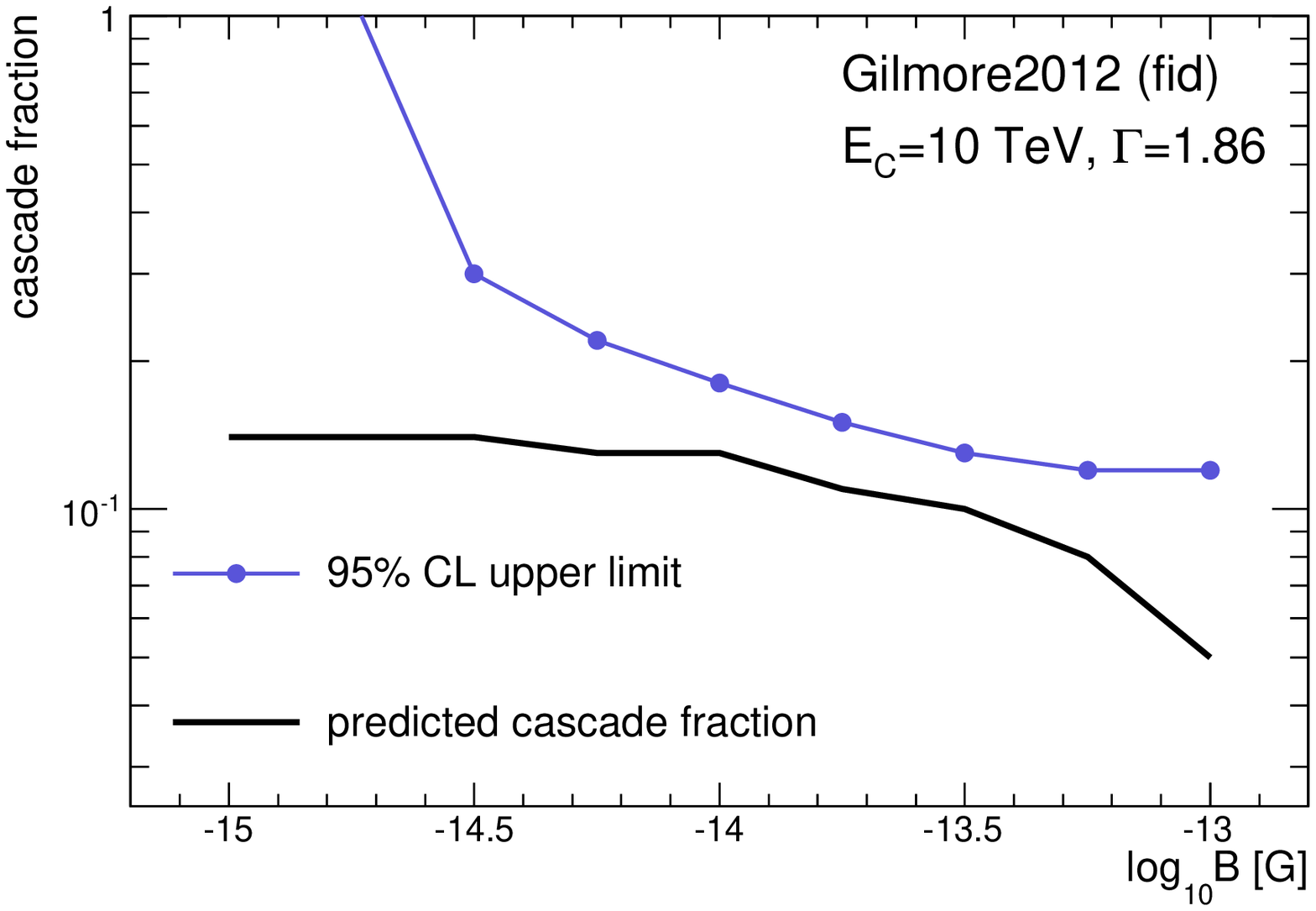}
\includegraphics[width=0.45\textwidth]{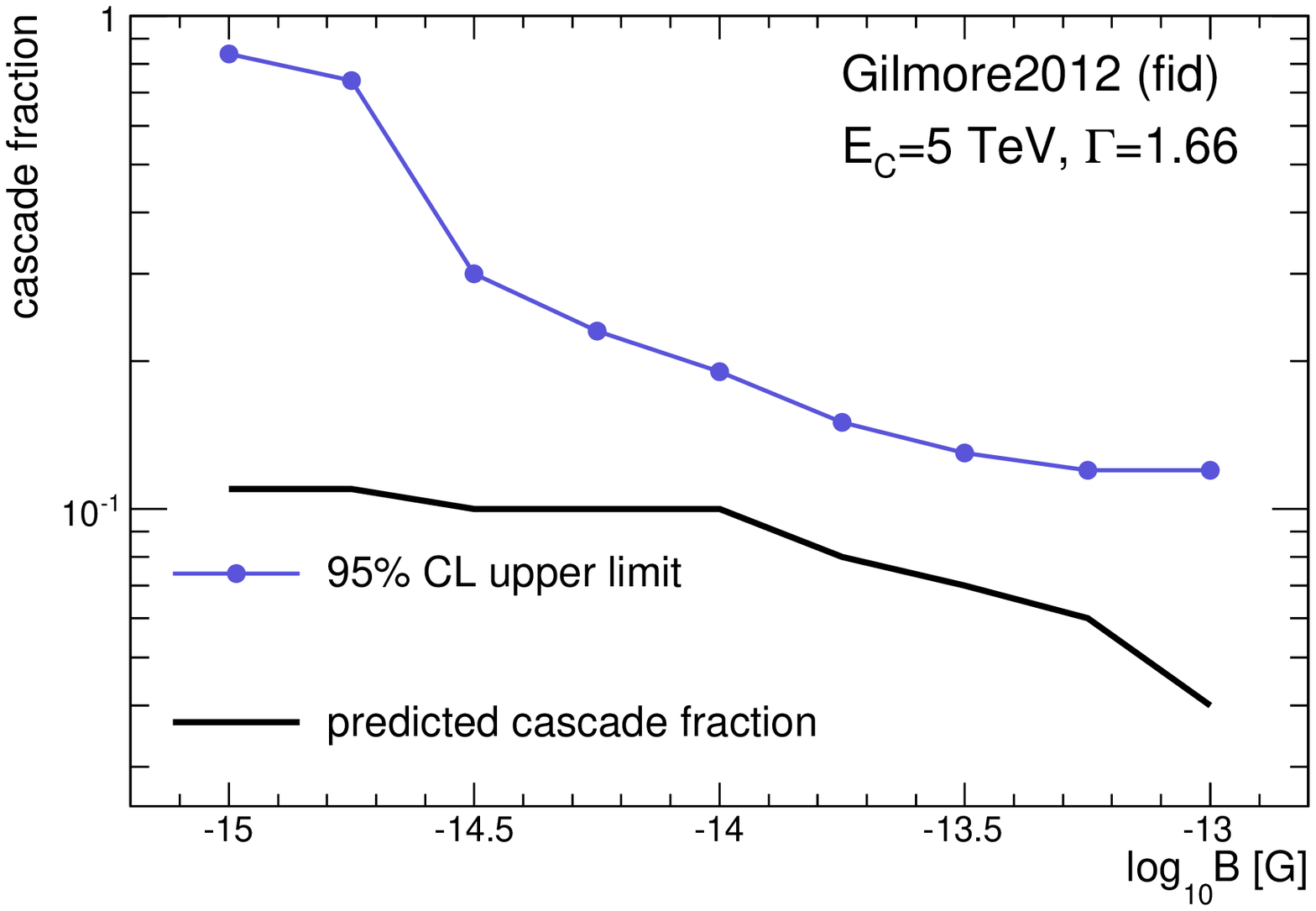}}
\centerline{\includegraphics[width=0.45\textwidth]{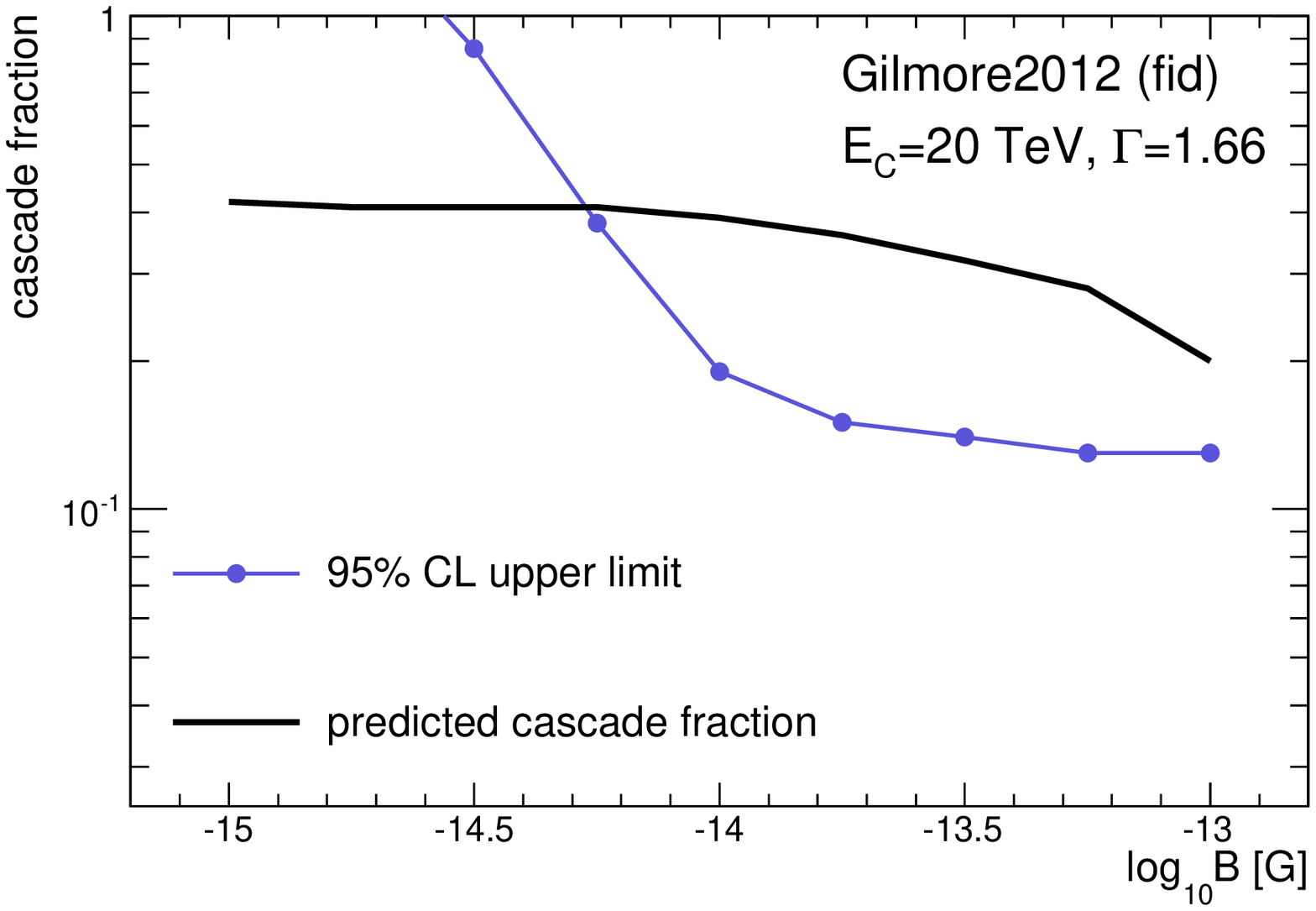}
\includegraphics[width=0.45\textwidth]{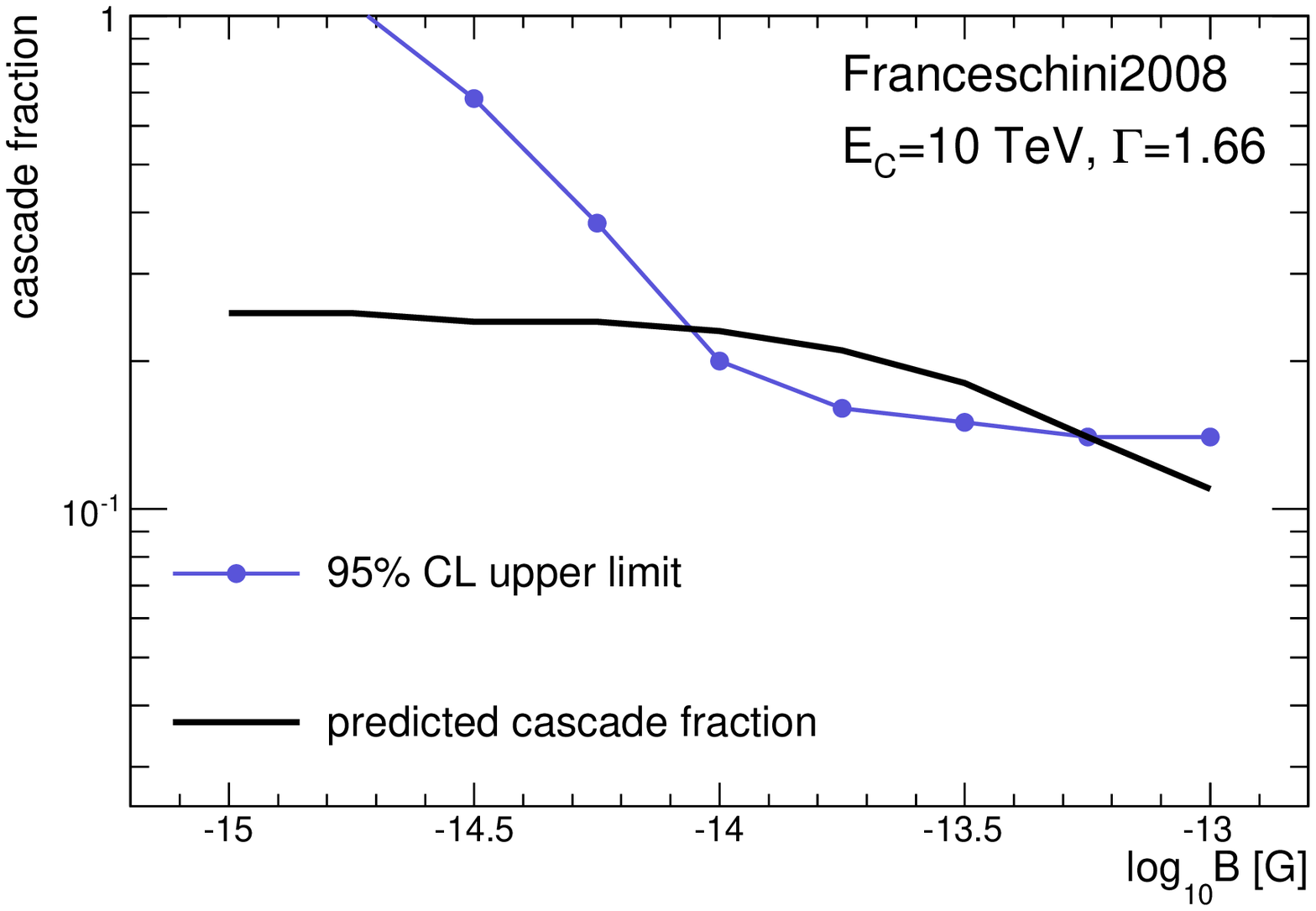}}
\caption{\small{The 95\% CL upper limits on the cascade fraction $f_{c}$ as a function of IGMF strength, for different assumptions about the intrinsic spectrum of 1ES~1218+304 and for two different EBL models.}}\label{casfrac_pred_UL}
\end{figure}

The projected sensitivity to broadening of the source angular distribution due to a cascade emission component hinges heavily on the intrinsic spectrum of the source. Based on the cascade simulations, the predicted cascade fraction ($f_{c}$; ratio of cascade emission to total emission) must be $\gtrsim$~10\% to produce an angular broadening that exceeds the statistical and systematic uncertainties on the widths in the datasets studied here. Evidence for an intrinsic cutoff below several TeV leads to a predicted $f_{c}$ of less than 1\% for all sources but the extreme-HBLs 1ES~0229+200 and 1ES~1218+304. The source 1ES 0229+200, although not showing evidence of an intrinsic cutoff, has the softest spectral index of the sources studied, at 2.025$\pm$0.150 in the HE range~\citep{3LAC}, which also results in a low predicted value of $f_{c}$. For 1ES 1218+304, the predicted $f_{c}$ is 10-25\% for the range of magnetic fields considered, as shown in Fig.~\ref{casfrac_pred_UL}. Consequently, of all the sources studied, only 1ES 1218+304 is used to place constraints on the IGMF strength. While a stacked analysis of all sources at similar redshifts was feasible, the combined limit would be entirely dominated by the contributions of 1ES 1218+304. Hence, a stacked analysis was not attempted.

However, several uncertainties on the predicted cascade emission remain, and their impact must be examined when deriving a limit on the IGMF strength.
\begin{itemize}
\item{Intrinsic cutoff; a cutoff at energies above the highest energy VERITAS spectral point cannot be excluded. Limits on the IGMF strength were derived assuming an exponential cutoff in the intrinsic spectrum at several energies: $E_{C}$=5, 10, and 20 TeV.}
\item{Spectral variability; the assumed value of the intrinsic spectral index of 1ES 1218+304, $\Gamma$=1.660~\citep{3LAC} is measured from the full \textit{Fermi}-LAT dataset. However, this does not account for any spectral variability that occurred either within this dataset or over the lifetime of the blazar (this is relevant as the cascade emission can, for a high IGMF strength, experience a time delay longer than the time for which VERITAS has been operating). The dependence of the IGMF limits on $\Gamma$ was tested by assuming $\Gamma$=1.460 and $\Gamma$=1.860, while fixing the cutoff energy to 10~TeV.}
\item{EBL model; the development of the cascade depends on the photon density predicted by the input EBL model. Limits on the IGMF strength were nominally derived assuming the fiducial model of \cite{Gilmore2012}. To estimate the sensitivity of the IGMF constraints to the EBL model, limits were also derived with the model of \cite{Franceschini2008}, while keeping $\Gamma$=1.660 and $E_{C}$=10~TeV fixed. These models were selected for their consistency with the EBL measurement of \cite{Biteau2015}.} 
\end{itemize}

For each assumed IGMF strength and set of model assumptions, a function describing the total (cascade and primary) emission was produced for 100 values of $f_{c}$ between 0 and 1. The $\theta^{2}$ function for the cascade emission was derived by convolving the simulated cascade emission's $\theta^{2}$ distribution with the PSF measured from the simulated point source for 1ES 1218+304. The total emission for a given $f_{c}$ is described by 
\begin{equation}
\textrm{total~emission}(\theta^{2}) = [(1 - f_{c})\times \textrm{primary~emission} + f_{c}\times \textrm{cascade~emission}],
\end{equation} 
where the function describing the primary emission is again taken from the simulated point source for 1ES~1218+304. For each value of $f_{c}$, an angular distribution for the total emission was simulated with $\sim$1 million events (matching the number of events for the simulated point source). The distributions were then fit with Equation~\ref{eq:Zitzerfunc}.

\begin{figure}
\centerline{\includegraphics[width=0.5\textwidth]{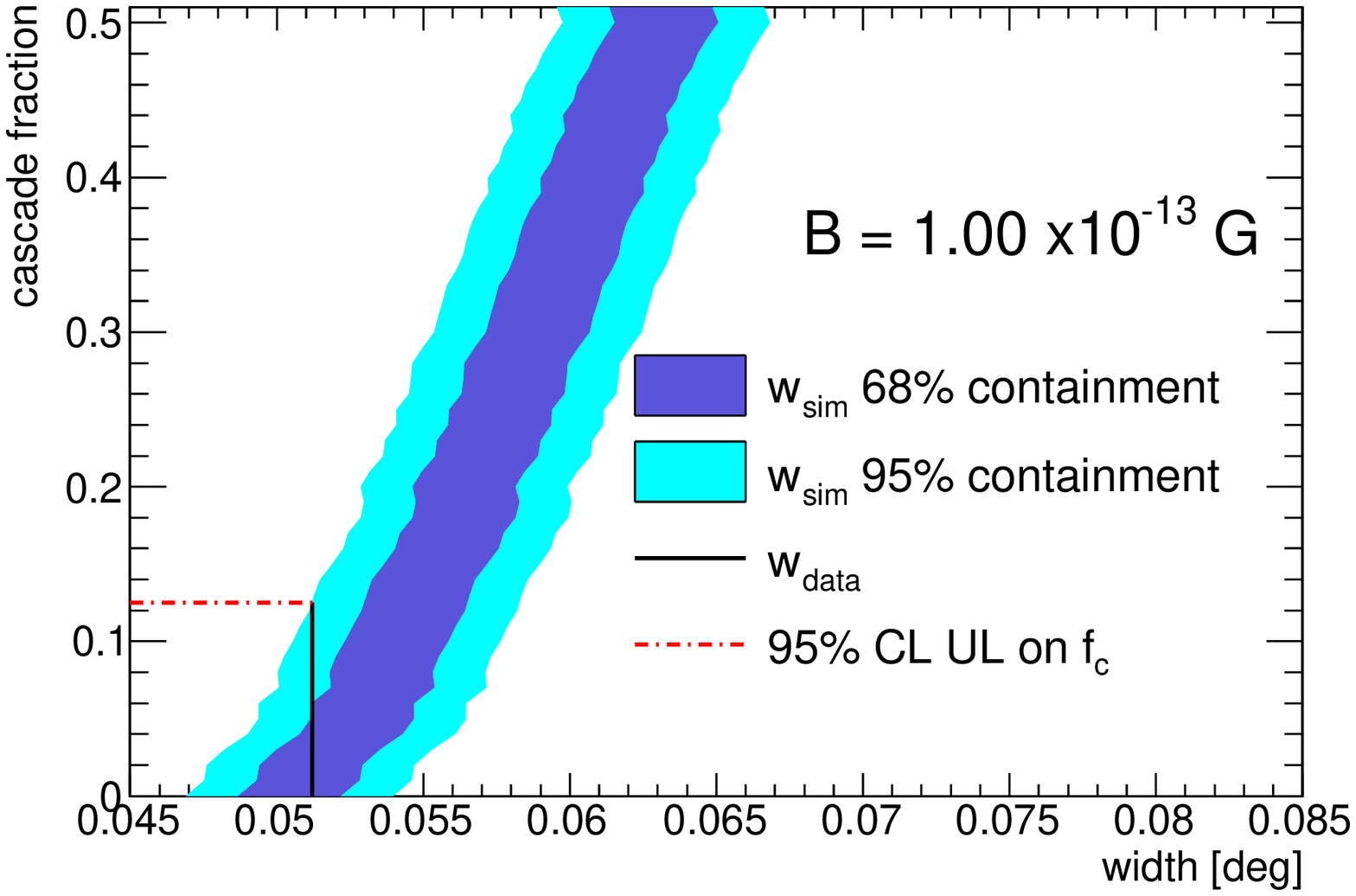}}
\caption{\small{The left panel shows the dependence of the width of the simulated angular distribution on the cascade fraction $f_{c}$ for 1ES 1218+304. This is compared against the width of the angular distribution measured in data, $w_{data}$.}}\label{width_v_fc_example}
\end{figure}

The left panel of Fig.~\ref{width_v_fc_example} shows the widths extracted from the fits versus $f_{c}$ for an IGMF strength of B = 10$^{-13}$~G, assuming the Gilmore 2012 fiducial model, $\Gamma$=1.660 and $E_{C}$=10~TeV. The uncertainty bands are given by the uncertainties on the simulated width shown in Table~\ref{fitwidths}, added in quadrature with the statistical uncertainty $\sigma_{stat}^{data}$. The measured width $w_{data}$ for 1ES~1218+304 is shown by the black vertical line. The red dashed horizontal line shows the upper limit on the cascade fraction; values of $f_{c}$ above this line are excluded at the 95\% confidence level (CL). 

For each set of model assumptions, the 95\% CL upper limit on $f_{c}$ as a function of IGMF strength was compared against the predicted cascade fraction obtained from the cascade simulations. The results are shown in Fig.~\ref{casfrac_pred_UL}. IGMF strengths for which the upper limit falls below the predicted $f_{c}$ are excluded at the 95\% CL. The exclusion ranges on the IGMF strength for each set of model assumptions are summarized in Table~\ref{IGMFlimits}. 

\begin{table}[!htp]
\centerline{
\begin{tabular}{cccc}
 $\Gamma$ & $E_{C}$ [TeV] &  EBL model & IGMF excluded [G] \\
\hline
\hline
1.660 & 10 & Gilmore2012 (fid) & 5.5$\times10^{-15}$--7.4$\times10^{-14}$  \\
1.460 & 10 & Gilmore2012 (fid) & 4.5$\times10^{-15}$--1.0$\times10^{-13}$  \\
1.860 & 10 & Gilmore2012 (fid) & non-constraining  \\
1.660 & 5   & Gilmore2012 (fid) & non-constraining   \\
1.660 & 20 & Gilmore2012 (fid) & 5.4$\times10^{-15}$--1.0$\times10^{-13}$   \\
1.660 & 10 & Francheschini2008 & 9.1$\times10^{-15}$--5.6$\times10^{-14}$ \\
\hline
\hline
\end{tabular}
}
\caption{\small{The 95\% confidence level exclusion ranges on the IGMF strength for each set of model assumptions.}}
\label{IGMFlimits}
\end{table}

\subsection{Limits on the Flux from Extended Emission}\label{fluxconstraints}
Upper limits on the integrated flux between 160~GeV and 1~TeV from angularly broadened emission are set for all sources. The bulk of the primary emission is expected to fall in the range $\theta^{2}$ = 0.00--0.01~deg$^{2}$, thus excess counts due to angularly broadened emission were calculated from the difference $\int \theta^{2}_{data} - \int \theta^{2}_{sim}$ within the integration range $\theta^{2}$=0.01--0.24~deg$^{2}$. The integration range was chosen to match the ranges used in similar calculations performed by \cite{HESS2014} and \cite{MAGIC2010}. Upper limits on the number of gamma-ray events due to angularly broadened emission are calculated using the frequentist method of Rolke~\citep{Rolke}, and translated into an upper limit on the rate by dividing by the deadtime-corrected exposure time. 

Translating the upper limit on the rate into an upper limit on the integrated flux requires an assumption about the spectral index of the angularly broadened emission. A spectral index ($\Gamma$+2)/2 was assumed, accounting for a slight softening of the cascade emission compared to the primary emission as inverse-Compton scattering proceeds in the Thompson limit. The resulting 95\% CL upper limits on the integrated flux due to angularly broadened emission for an energy range between 160~GeV and 1~TeV are shown in Table~\ref{modelindlimits}.

\begin{table}[!htp]
\centerline{
\begin{tabular}{cc}
Source name  &  $95\%~CL$ [$10^{-12}~$cm$^{-2}$s$^{-1}$] \\
\hline
\hline
Mrk 421 & 1.4 \\
Mrk 501 & 4.2 \\
VER J0521+211 & 0.6 \\ 
H 1426+428 & 2.1 \\ 
1ES 0229+200 & 1.2 \\
1ES 1218+304 & 0.9 \\ 
PG 1553+113 & 3.2 \\ 
\hline
\hline
\end{tabular}
}
\caption{\small{Limits on the integrated flux from angularly broadened emission in an energy range between 160~GeV and 1~TeV, assuming a spectral index of ($\Gamma$+2)/2 for the cascade emission.}}
\label{modelindlimits}
\end{table}

\section{Discussion and Conclusions}
A search for source extension due to cascade emission broadened by the IGMF was performed with VERITAS observations of seven blazars. No indication of angularly broadened emission was observed. Limits were set on the fraction of the total emission due to cascade emission ($f_{c}$) for the blazar with the largest predicted cascade fraction, 1ES~1218+304. IGMF strengths between 10$^{-16}$ and 10$^{-13}$~G and an IGMF coherence length of 1 Mpc were assumed. Exclusion regions on the IGMF strength were determined under different sets of assumptions about the source intrinsic spectrum and the EBL intensity. For a nominal set of assumptions (spectral index $\Gamma$=1.660 and cutoff energy $E_{C}$=10~TeV for 1ES~1218+304, EBL model of \cite{Gilmore2012}), an IGMF strength of 5.5$\times10^{-15}$~G~--~7.4$\times10^{-14}$~G can be excluded at the 95\% CL. This shows a similar sensitivity to measurements from other instruments, as well as complementarity to previous results. Namely, H.E.S.S. ruled out an IGMF strength in the range (0.3--3)$\times$10$^{-15}$~G at the 99\% CL, using observations of PKS 2155-304 and slightly different model assumptions but otherwise similar methodology~\citep{HESS2014}. Taken together, the H.E.S.S. and VERITAS constraints rule out an IGMF strength falling in much of the range between 10$^{-16}$ and 10$^{-13}$ G. The VERITAS exclusion region, however, does not rule out the IGMF range suggested by the claimed detection of \cite{Chen2015}.

Varying the assumptions on the intrinsic spectrum of the source, namely the spectral index and the high-energy cutoff, substantially alters the extracted limits on $f_{c}$. Softening the assumed intrinsic spectral index by 0.2 or decreasing the energy of the exponential cutoff from 10 TeV to 5 TeV resulted in limits on $f_{c}$ falling above the predicted cascade fraction. In these cases, the IGMF strength is not constrained. The assumed shape of the cutoff impacts the constraints as well: assuming a super exponential cutoff power law exp($-(E/E_{C})^{\gamma}$) will produce stronger constraints for 0 $<\gamma<$ 1 (softer cutoff) and weaker constraints for $\gamma>$ 1 (sharper cutoff) than for the assumed exponential cutoff power law ($\gamma$=1). Finally, the EBL model assumed when simulating the cascade process affects the limits. It was observed that using the model of \cite{Gilmore2012} in the cascade simulations produced a broader IGMF exclusion region than using the model of \cite{Franceschini2008}. 

As the cascade emission is time delayed by years for the IGMF strengths considered here, the flux variability of the source over its lifetime will impact the limits on the IGMF strength, as the predicted $f_{c}$ is derived assuming the currently observed flux of primary emission. If the source exhibited a lower (higher) flux at the time that the cascade emission reaching the observer today was produced, $f_{c}$ will be lower (higher) than expected based on the current flux of 1ES~1218+308. Based on the ratio of the observed upper limits on $f_{c}$ and the predicted values, the differential flux of 1ES~1218+308 at 1~TeV would have to be $\sim$70\% of its current value on average over its lifetime to invalidate the entire nominal IGMF exclusion range.
 
It should be noted that while the assumed intrinsic source spectrum and EBL model affect the IGMF limits, the results are not expected to be sensitive to the choice of Doppler factor or viewing angle of the jet, nor to the choice of correlation length. It was demonstrated in \cite{Arlen2014} that the cascade spectrum above 100~GeV is unaffected by variation of the bulk Lorentz factor between 5 and 100, or variation of the jet viewing angle between 0$^{\circ}$ and 10$^{\circ}$. The IGMF limits are insensitive to the correlation length provided that the correlation length exceeds the inverse-Compton cooling length \citep{Neronov2009}. The cooling length and primary gamma-ray energy scale inversely. The primary gamma rays must have energies above 1 TeV to produce cascade emission with energies above the VERITAS energy threshold. Thus, the majority of the primary gamma rays will have cooling lengths of less than a few hundred kpc, much less than the correlation length of 1~Mpc used in the cascade simulation code.
 
For the cutoff energy of 10 TeV assumed for the intrinsic spectrum of 1ES 1218+308, the first pair production interaction occurs $>$ 10 Mpc from the source. Consequently, this study probes the magnetic field strength in areas distant from the source, sampling cosmic voids, rather than matter-rich regions.
 
Limits were set on the integrated flux due to angularly broadened emission for all sources, resulting in 95\% CL upper limits of (0.6--4.2)$\times10^{-12}$~cm$^{-2}$s$^{-1}$ for an energy range between 160~GeV and 1~TeV. A spectral index of ($\Gamma$+2)/2 was assumed for the angularly broadened emission.
 
The IGMF constraints presented here are limited by a number of factors, however the dominant limitation on the sensitivity to angularly broadened emission is the instrument point spread function. The Cherenkov Telescope Array (CTA) is projected to begin taking data in several years with a sensitivity greater than currently operating instruments, and in particular, with a substantially better PSF compared to VERITAS~\citep{Meyer2016, CTAconcept}, and consequently an improved ability to probe weaker IGMF strengths and smaller cascade fractions.



\acknowledgments

This research is supported by grants from the U.S. Department of Energy Office of Science, the U.S. National Science Foundation and the Smithsonian Institution, and by NSERC in Canada. E. Pueschel acknowledges the support of a Marie Curie Intra-European Fellowship within the 7th European Community Framework Programme, and thanks Andrew Taylor for useful discussion. We acknowledge the excellent work of the technical support staff at the Fred Lawrence Whipple Observatory and at the collaborating institutions in the construction and operation of the instrument. The VERITAS Collaboration is grateful to Trevor Weekes for his seminal contributions and leadership in the field of VHE gamma-ray astrophysics, which made this study possible. 



\clearpage
\appendix
\section{Angular profile comparison for sources and simulated point sources}\label{appendixA}
\begin{figure}[!htbp]
\centerline{\includegraphics[width=0.33\textwidth]{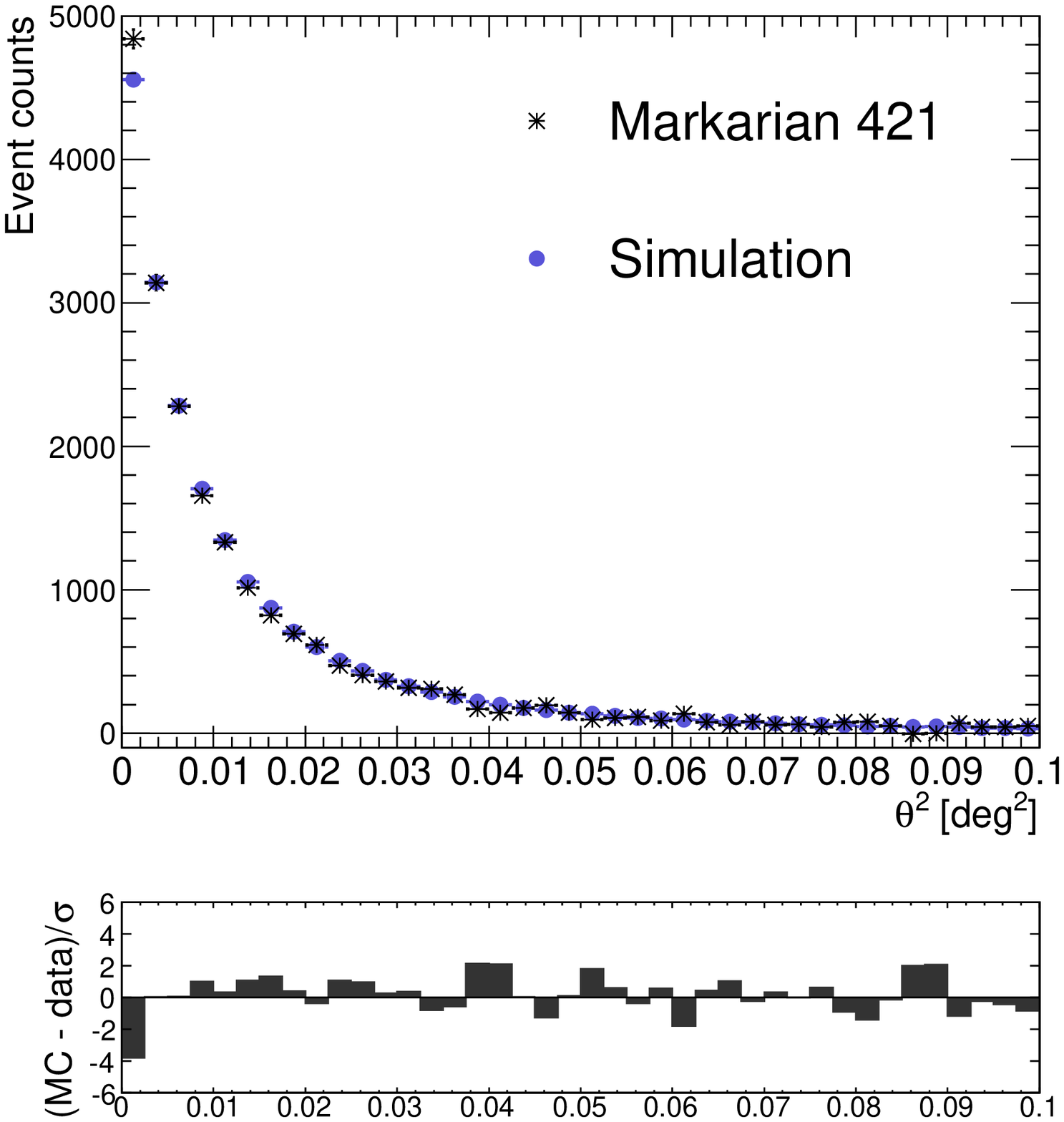}
\includegraphics[width=0.33\textwidth]{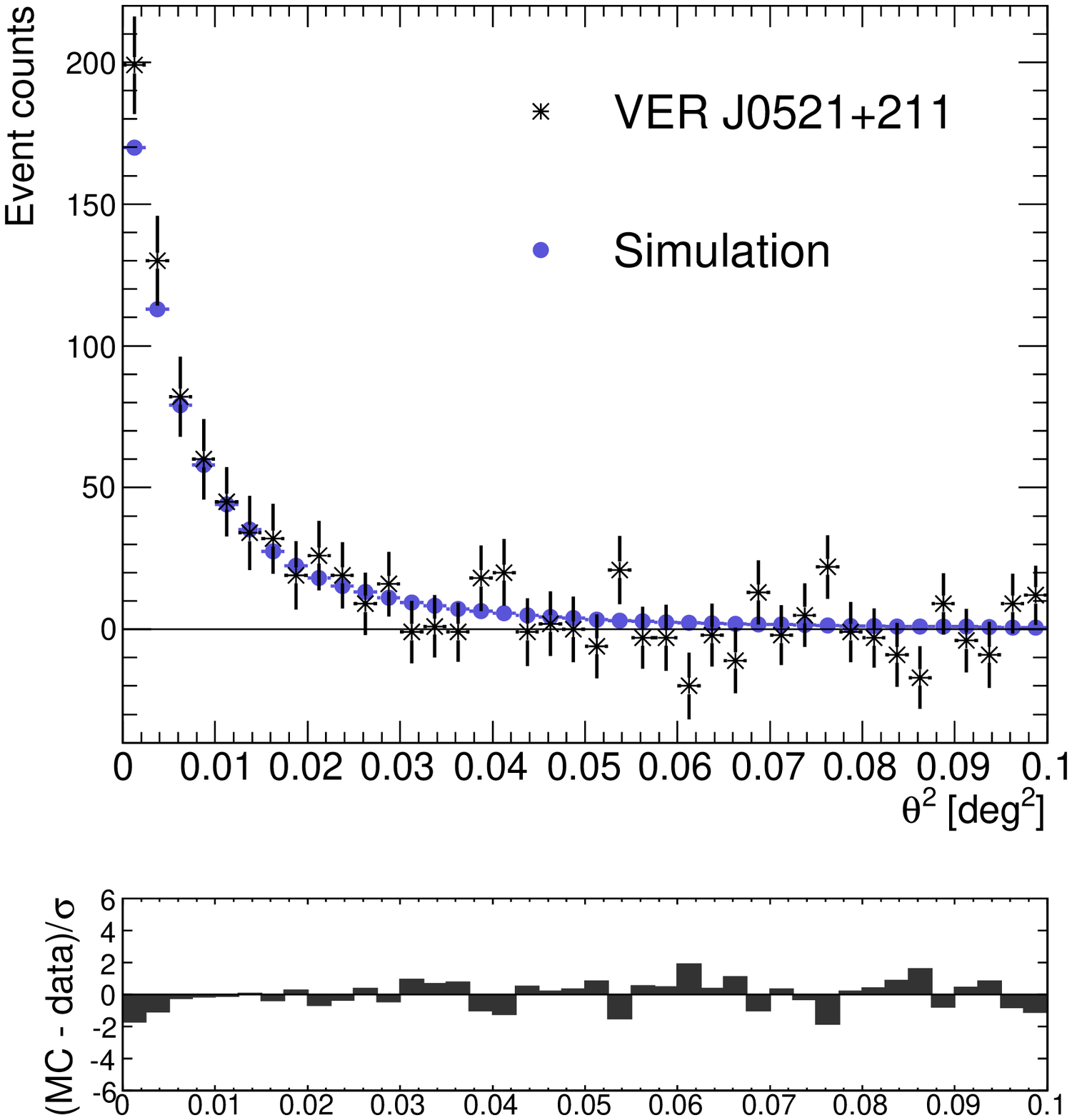}
\includegraphics[width=0.33\textwidth]{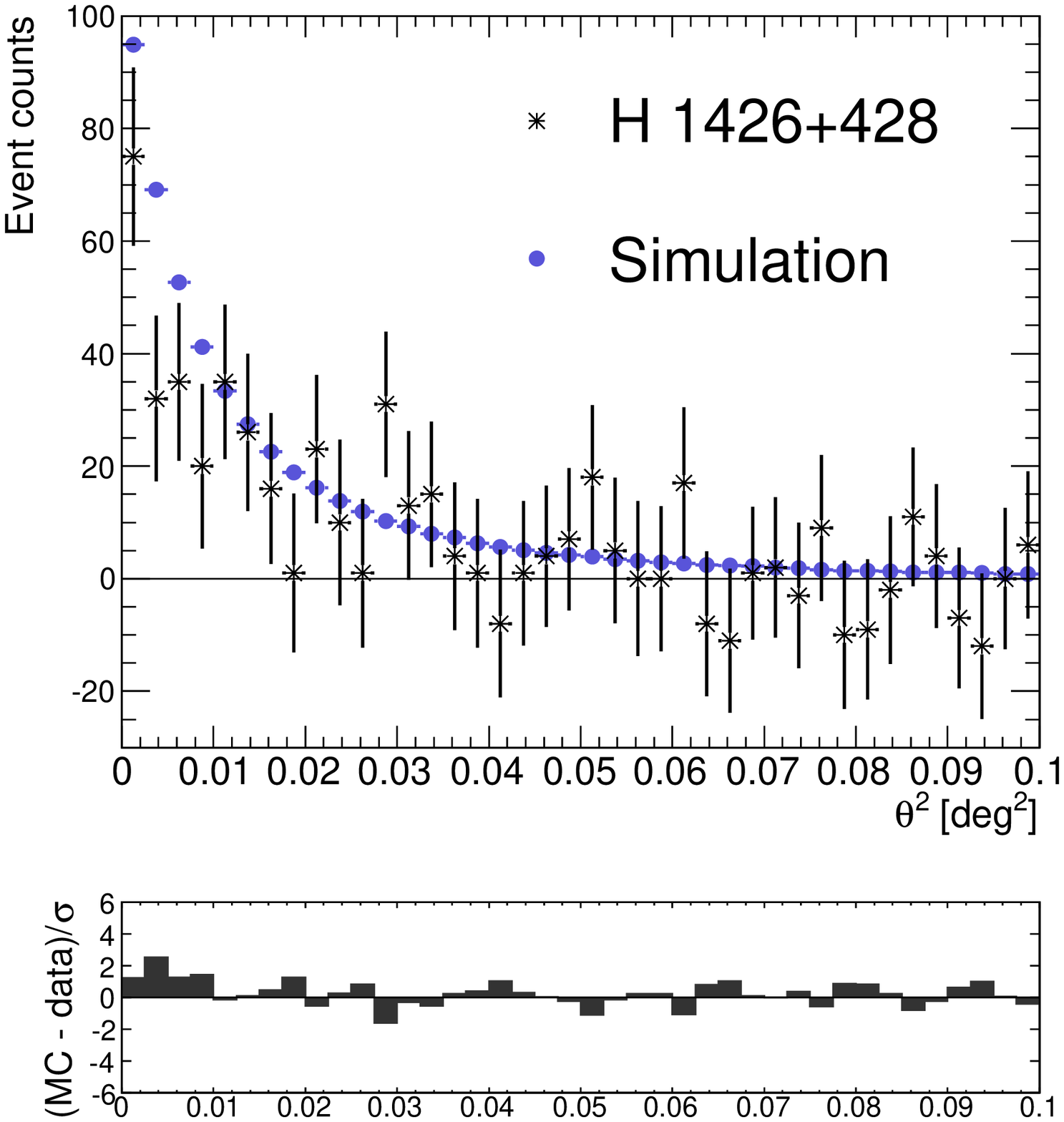}}
\centerline{\includegraphics[width=0.33\textwidth]{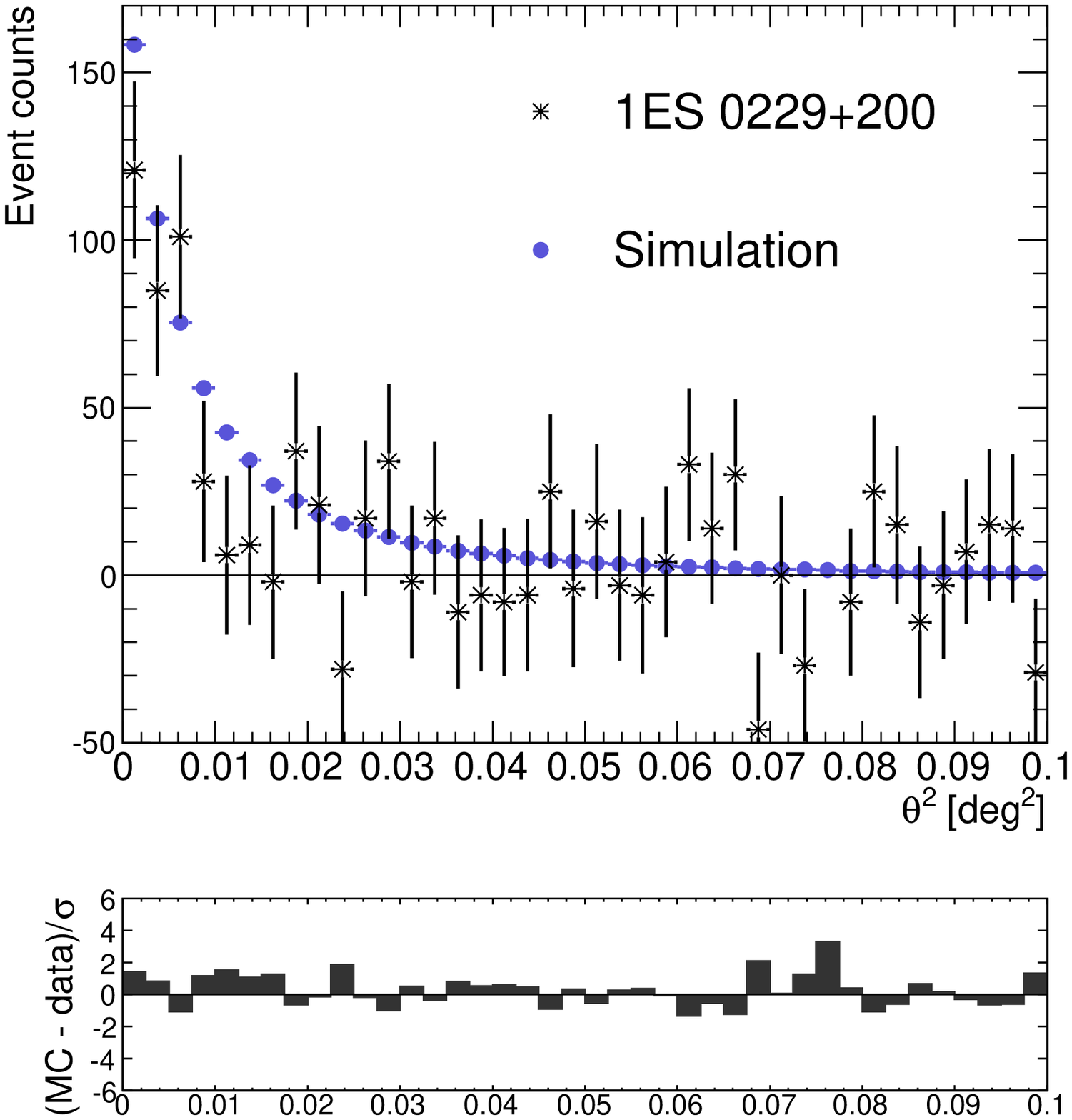}
\includegraphics[width=0.33\textwidth]{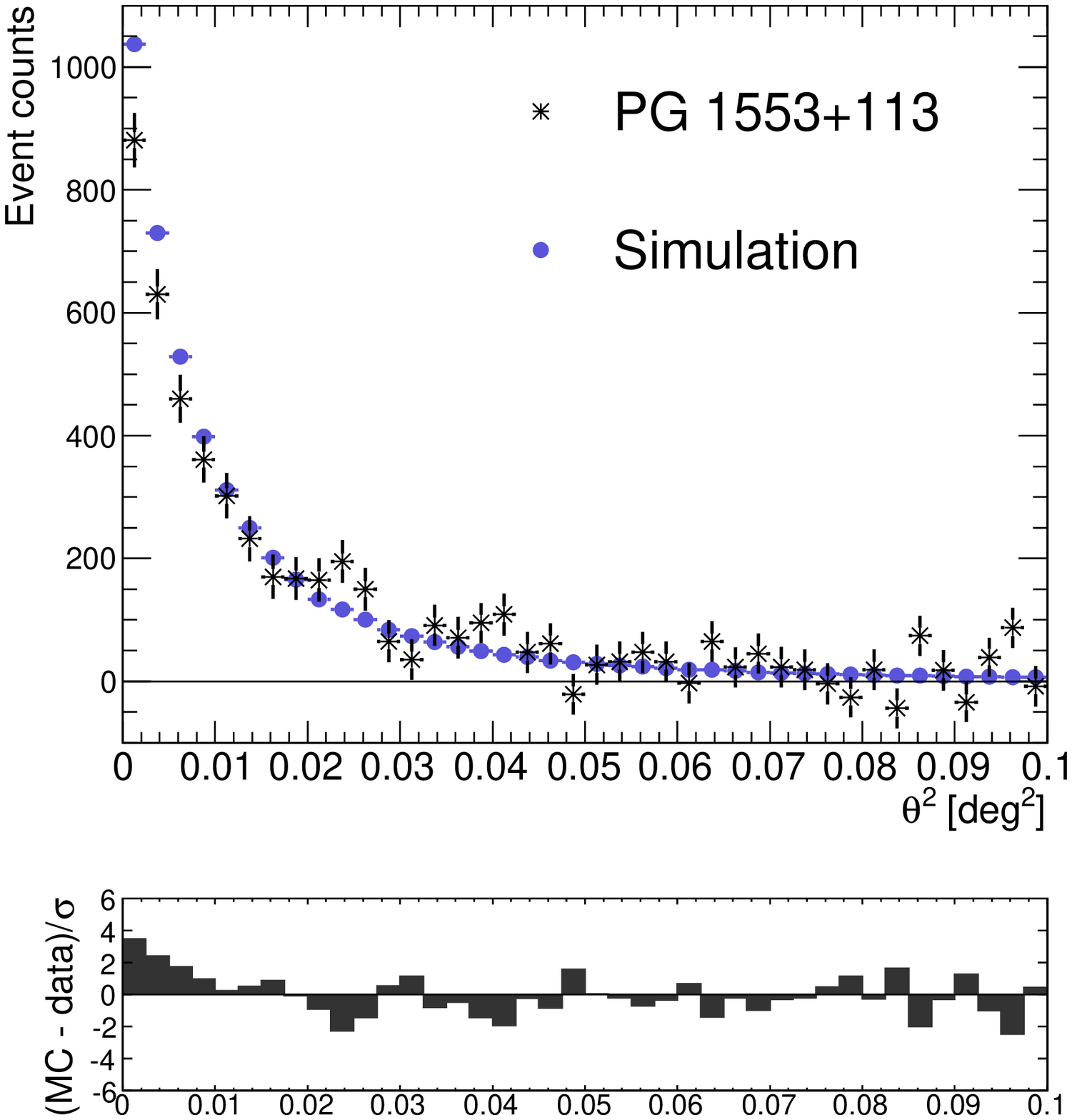}}
\caption{\small{Comparison between the angular profiles of the observed sources and their simulated counterparts. The results of a $\chi^{2}$ probability test are shown in Table~\protect\ref{sourcelist}.}}
\end{figure}

\clearpage
\section{Fitted angular profiles for sources and simulated point sources}\label{appendixB}
\begin{figure}[htbp]
\centerline{\includegraphics[width=0.4\textwidth]{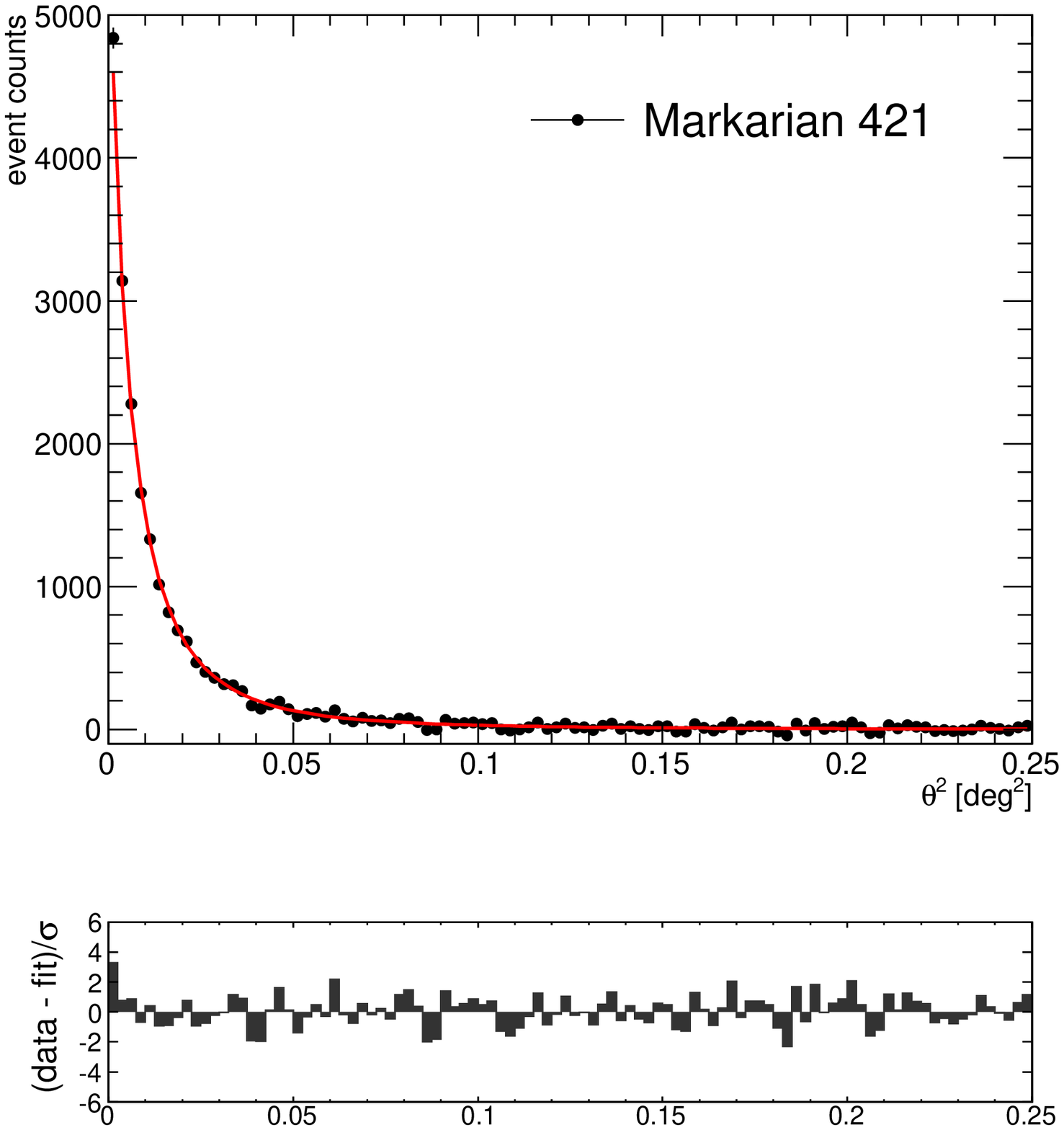}
\includegraphics[width=0.4\textwidth]{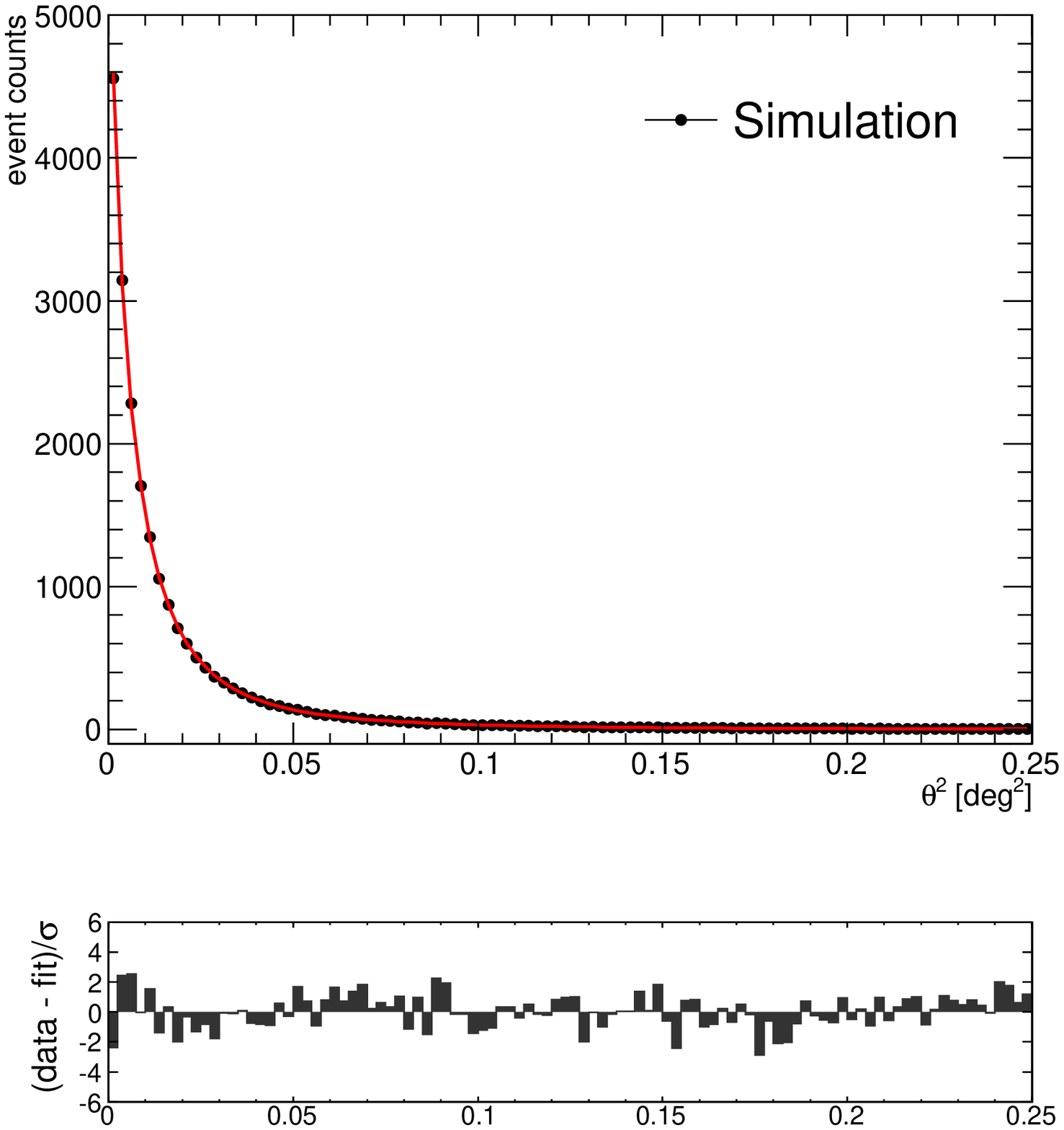}}
\caption{\small{Fitted $\theta^{2}$ distribution for Mrk 421 and its simulated counterpart.}}\label{fitMrk421}
\end{figure}

\begin{figure}[htbp]
\centerline{\includegraphics[width=0.4\textwidth]{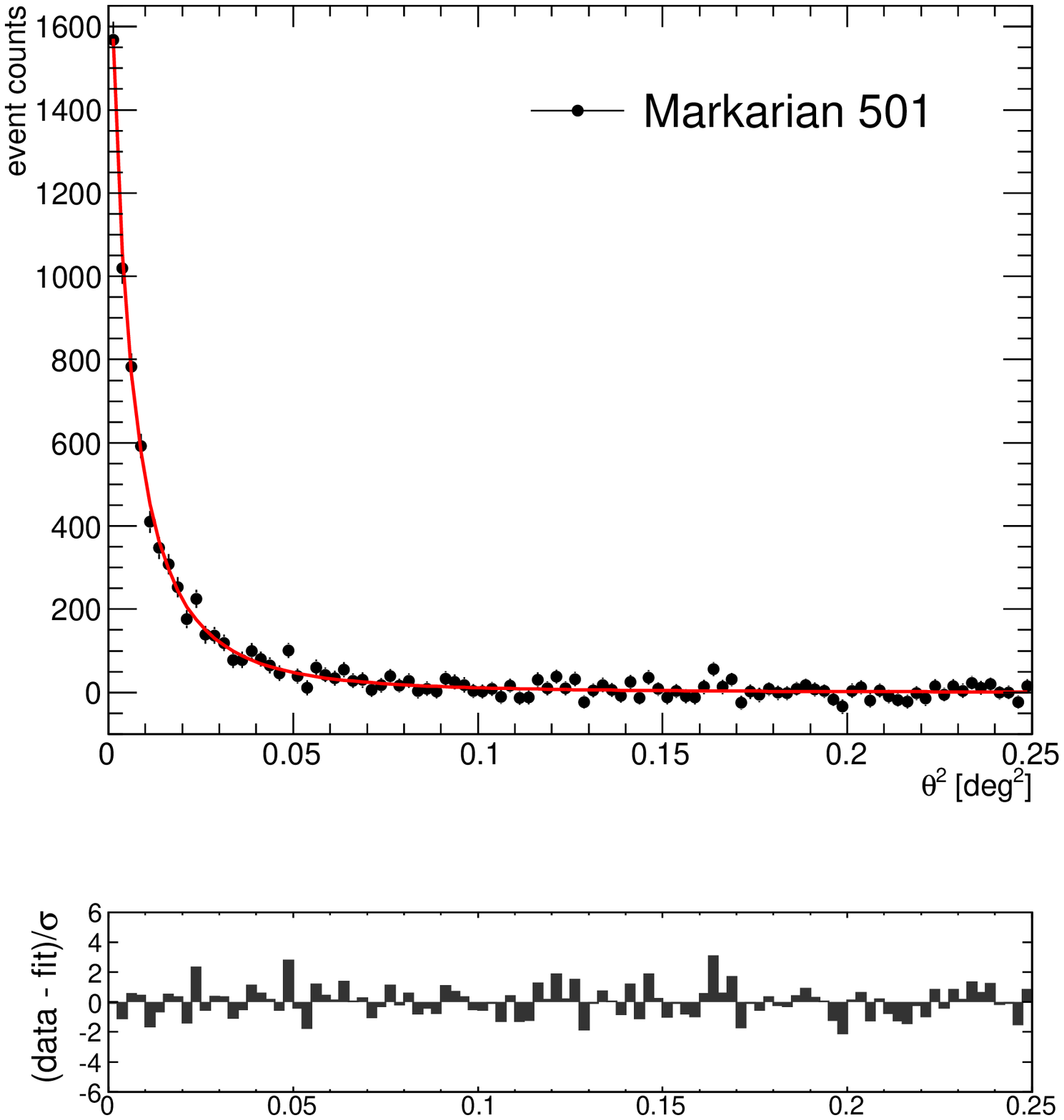}
\includegraphics[width=0.4\textwidth]{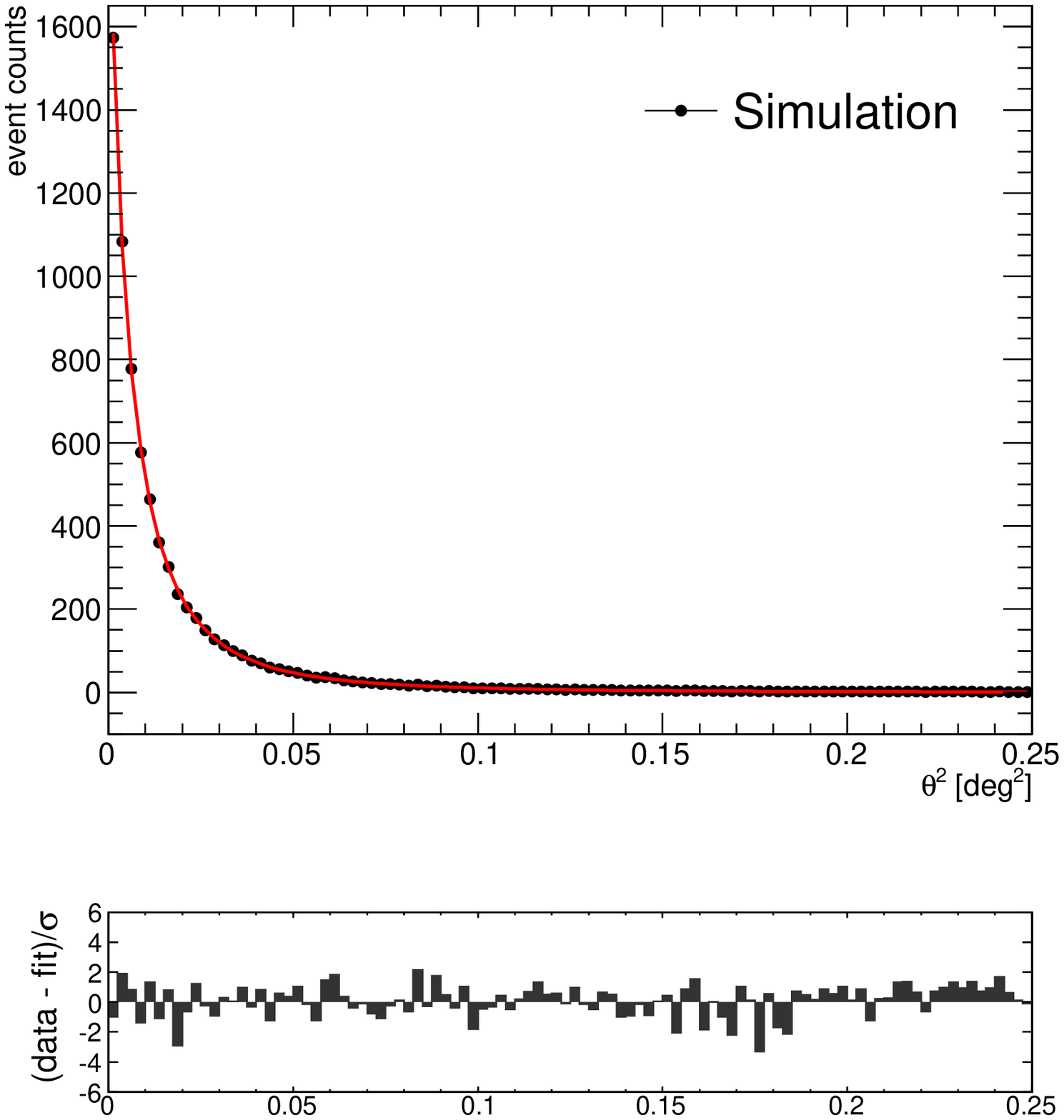}}
\caption{\small{Fitted $\theta^{2}$ distribution for Mrk 501 and its simulated counterpart.}}\label{fitMrk501}
\end{figure}

\begin{figure}[htbp]
\centerline{\includegraphics[width=0.4\textwidth]{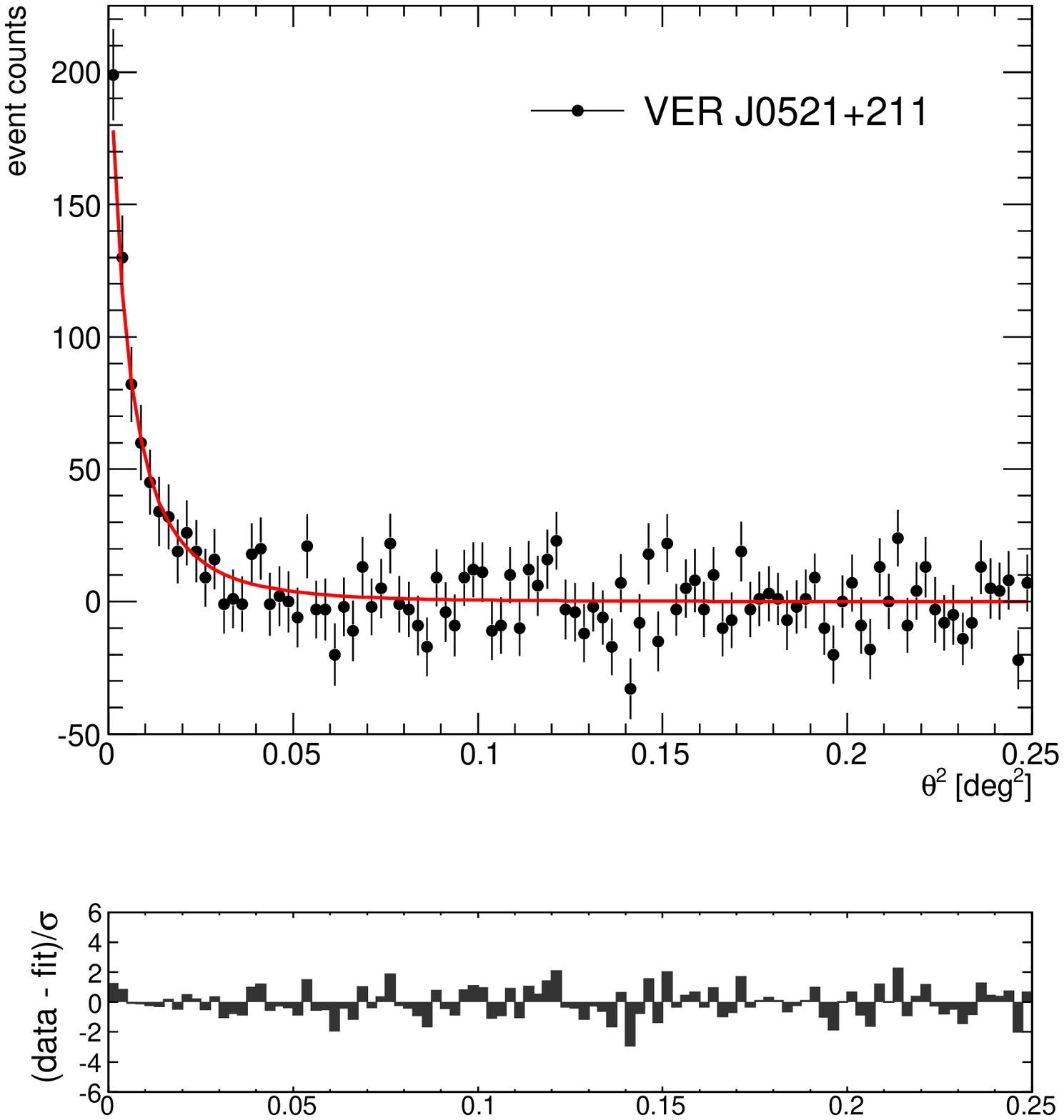}
\includegraphics[width=0.4\textwidth]{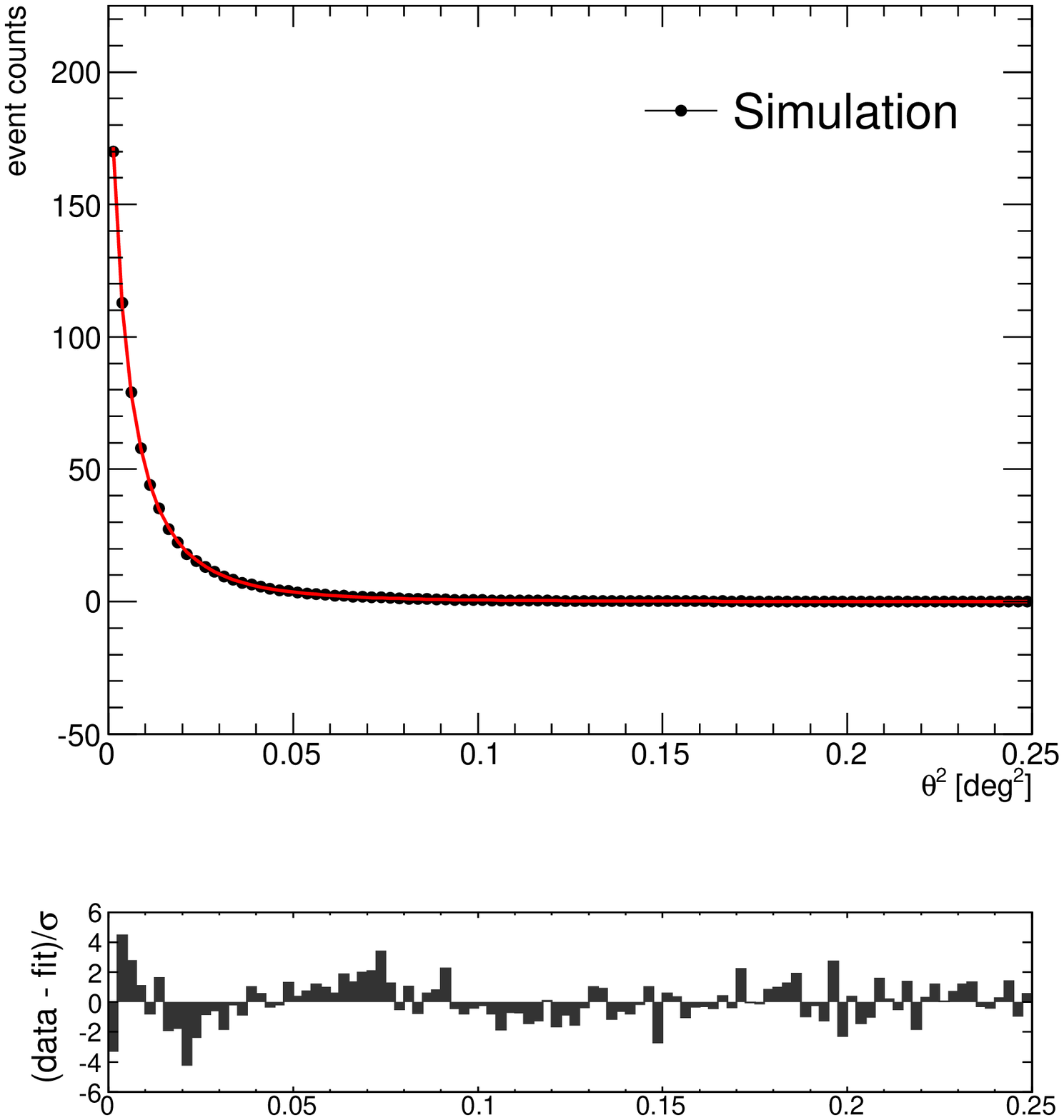}}
\caption{\small{Fitted $\theta^{2}$ distribution for VER J0521+211 and its simulated counterpart.}}\label{fitVERJ0521}
\end{figure}

\begin{figure}[htbp]
\centerline{\includegraphics[width=0.4\textwidth]{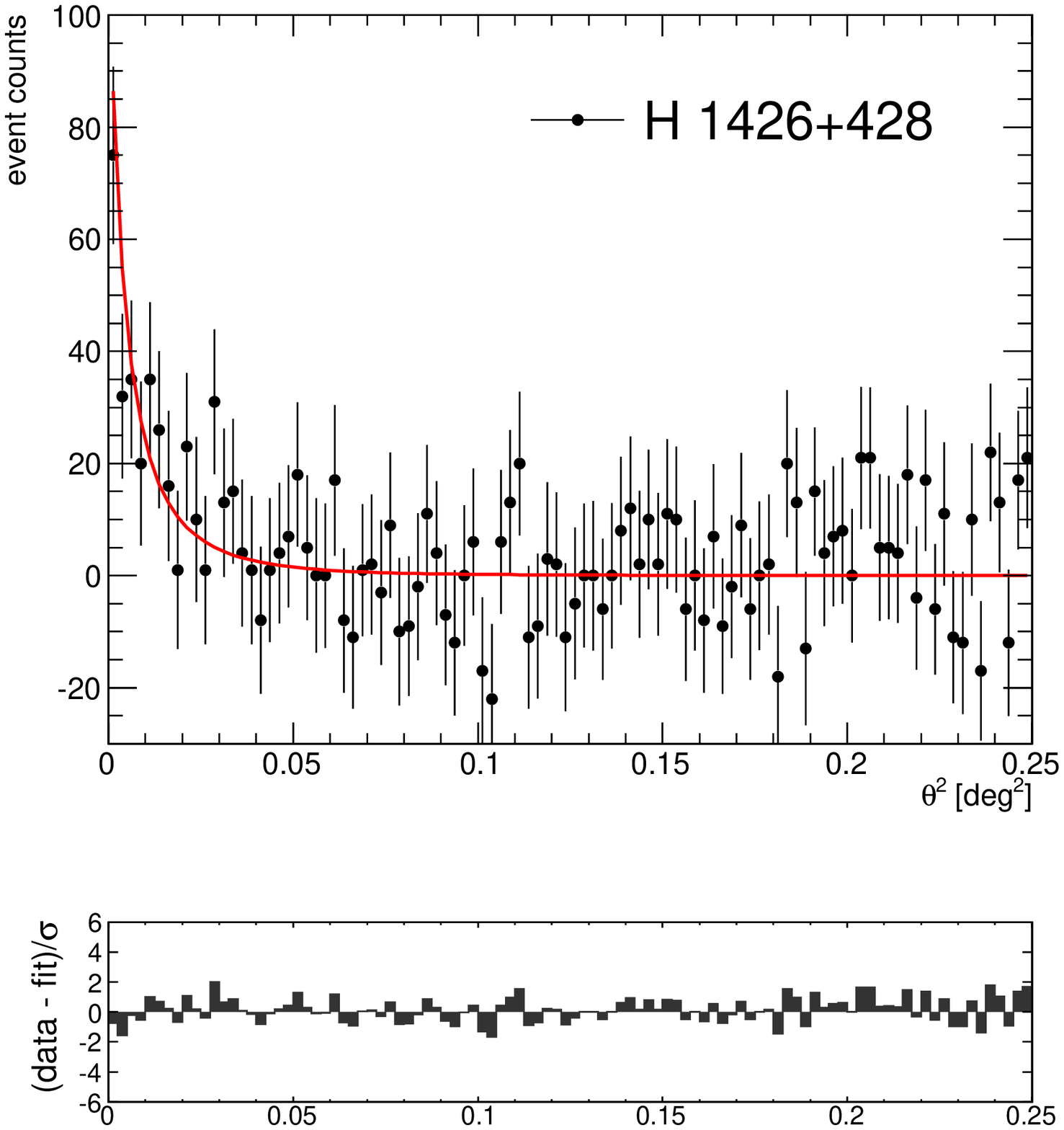}
\includegraphics[width=0.4\textwidth]{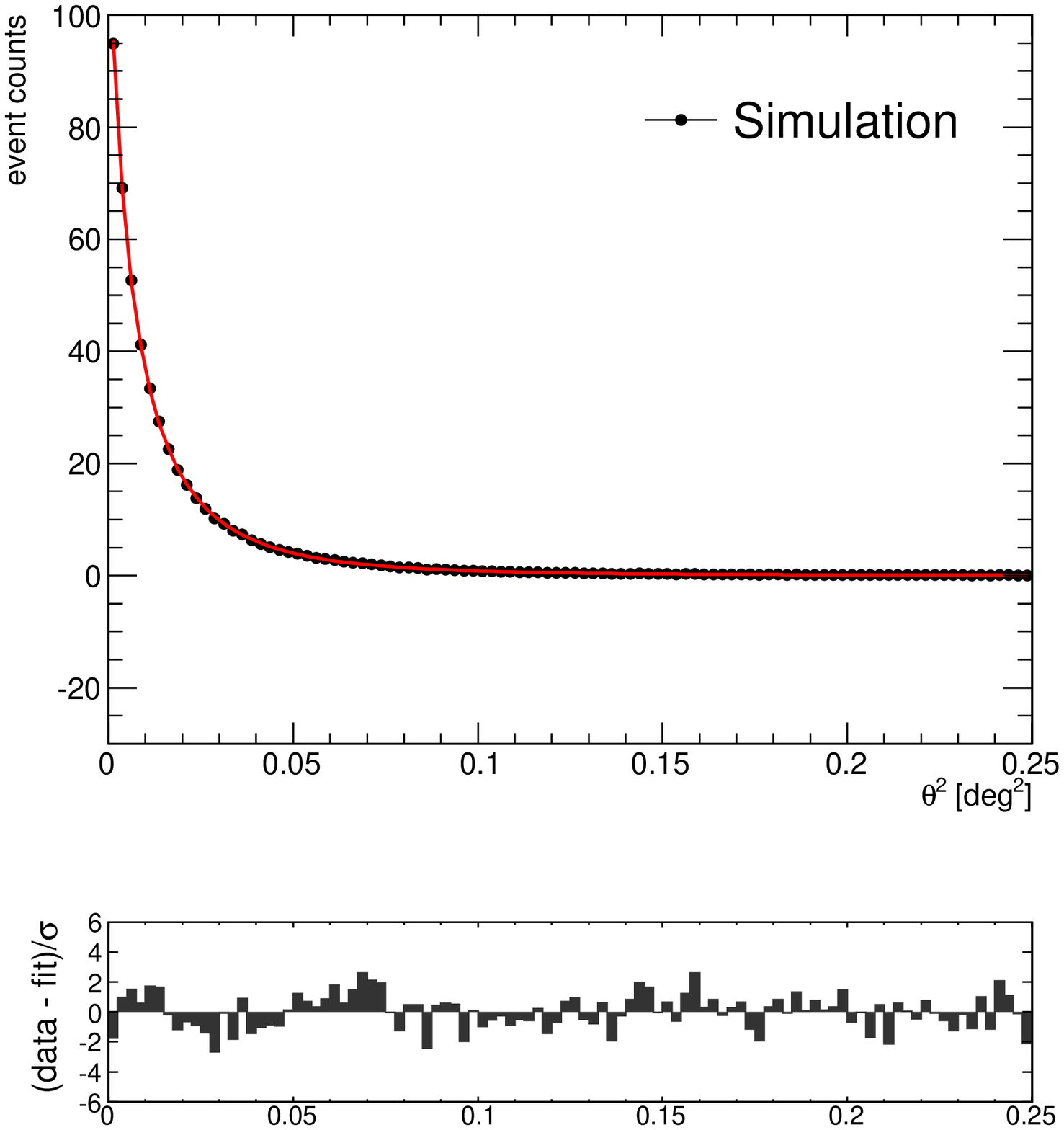}}
\caption{\small{Fitted $\theta^{2}$ distribution for H 1426+428 and its simulated counterpart.}}\label{fitH1426}
\end{figure}

\begin{figure}[htbp]
\centerline{\includegraphics[width=0.4\textwidth]{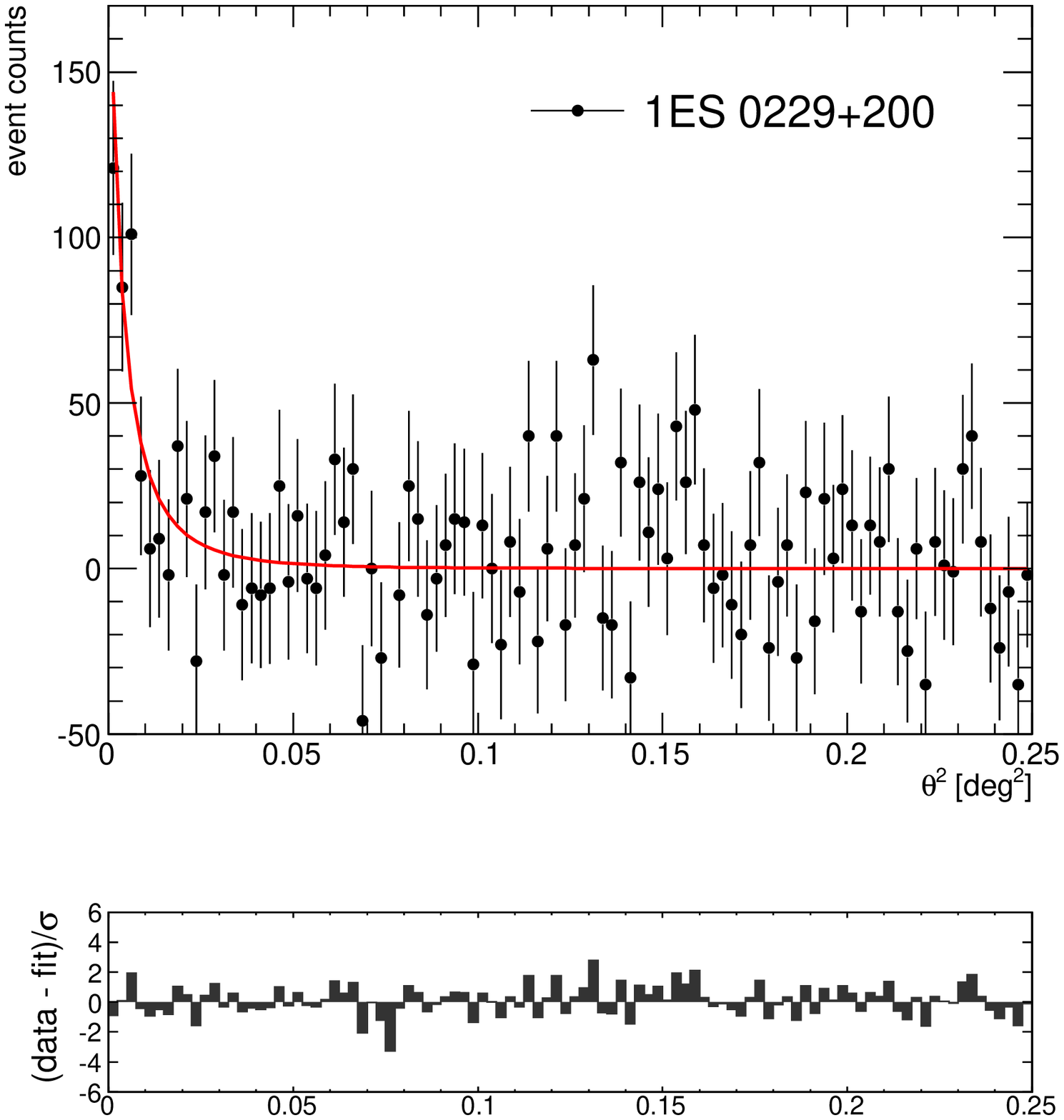}
\includegraphics[width=0.4\textwidth]{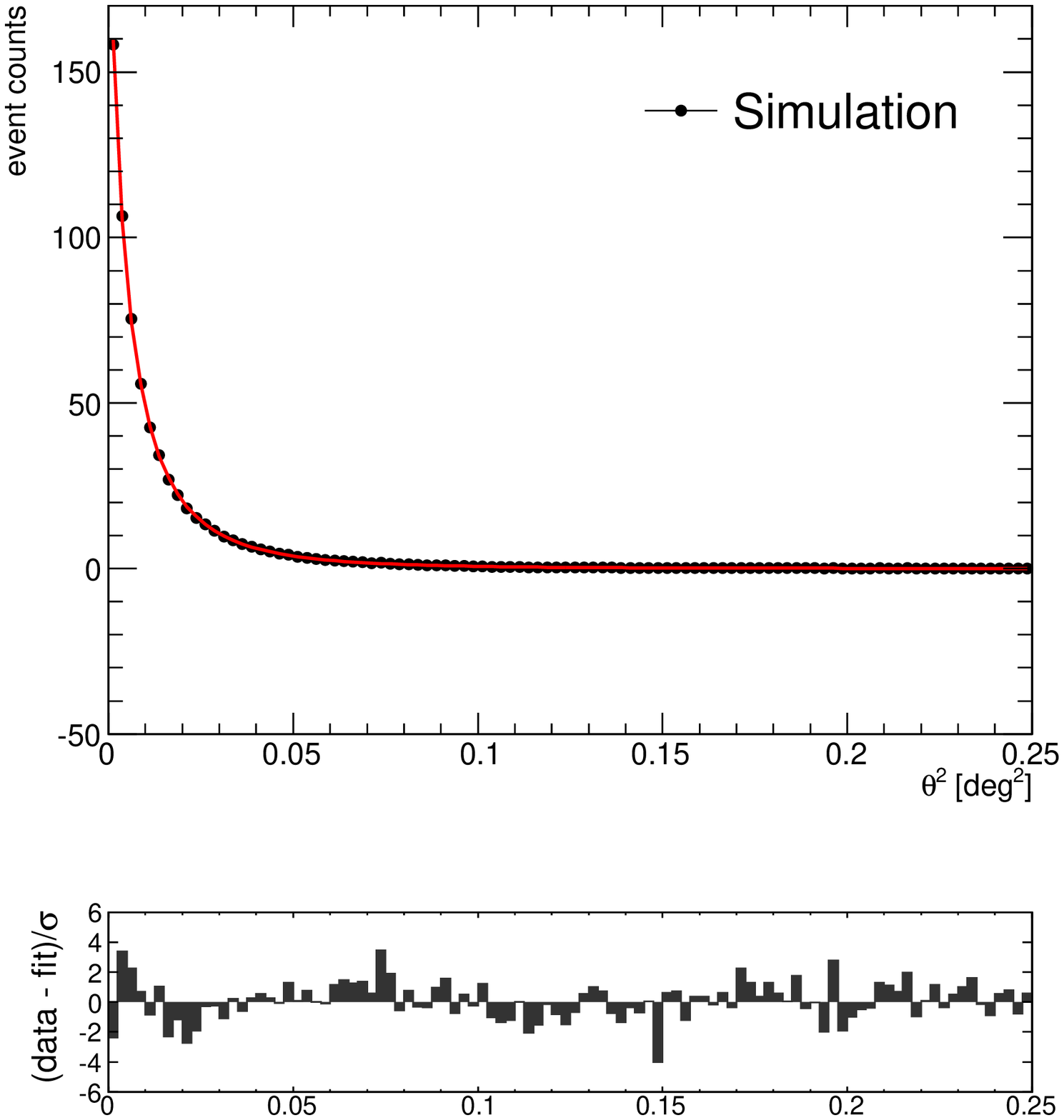}}
\caption{\small{Fitted $\theta^{2}$ distribution for 1ES 0229+200 and its simulated counterpart.}}\label{fit1ES0229}
\end{figure}

\begin{figure}[htbp]
\centerline{\includegraphics[width=0.4\textwidth]{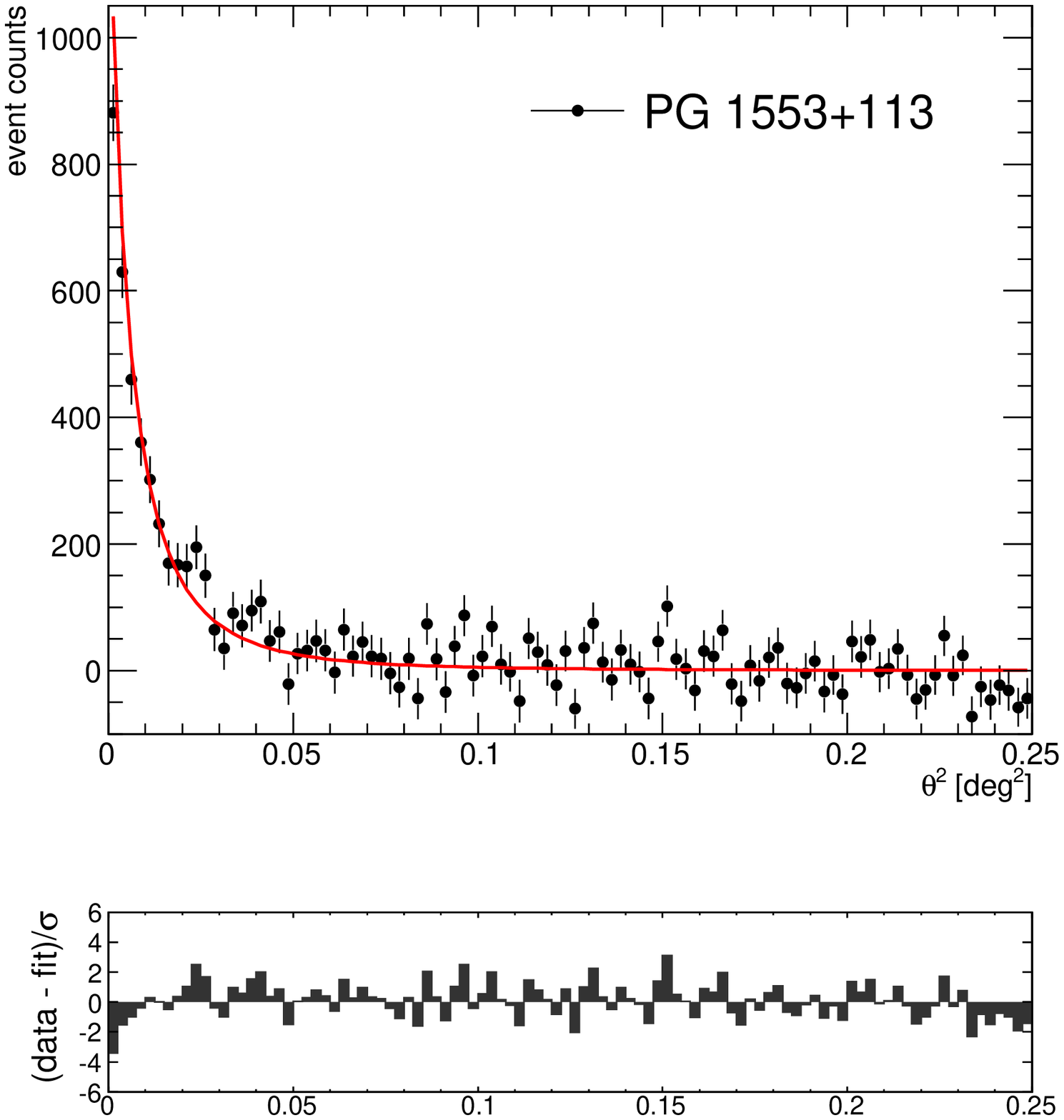}
\includegraphics[width=0.4\textwidth]{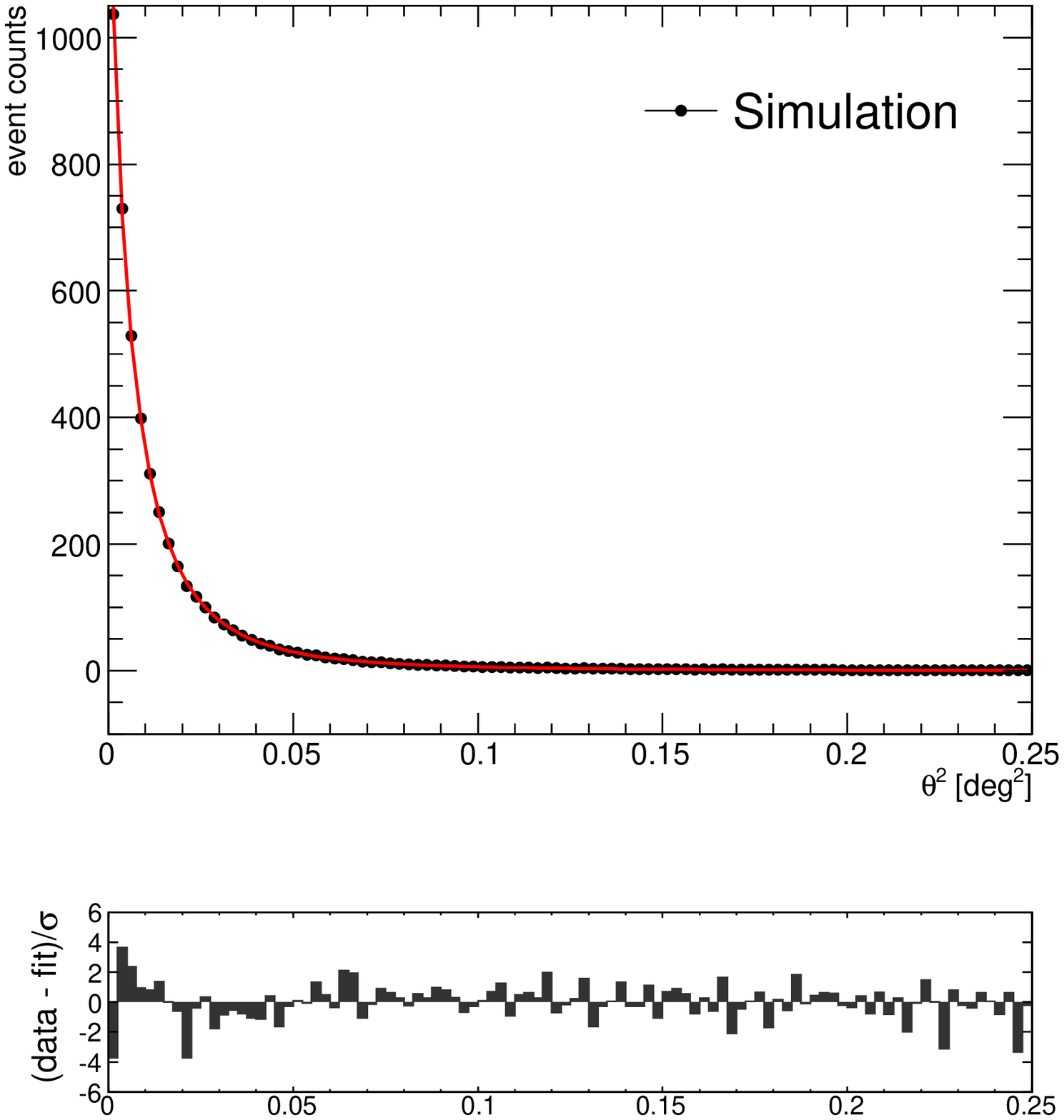}}
\caption{\small{Fitted $\theta^{2}$ distribution for PG 1553+113 and its simulated counterpart.}}\label{fitPG1553}
\end{figure}

\clearpage



\clearpage



\end{document}